\documentclass[article,showkeys,showpacs,preprintnumbers,amsmath,amssymb,aps,nofootinbib]{revtex4-1}
\usepackage{graphicx}
\usepackage{dcolumn}
\usepackage{bm}
\usepackage{xcolor}
\usepackage{bbm}
\usepackage{amssymb}
\usepackage{amsmath}
\usepackage{textcomp}
\usepackage[bottom]{footmisc}
\usepackage{wrapfig}
\usepackage{tikz}
\usetikzlibrary{scopes,intersections}
\usepackage{pgfplots}
\usepackage{pst-solides3d}
\usetikzlibrary{arrows,positioning}
\usetikzlibrary{decorations.pathmorphing}
\tikzset{zigzag/.style={decorate, decoration=zigzag}}

\makeatletter
\def\@hex@@Hex#1%
 {\if a#1A\else \if b#1B\else \if c#1C\else \if d#1D\else
  \if e#1E\else \if f#1F\else #1\fi\fi\fi\fi\fi\fi \@hex@Hex}
\makeatother
\definecolor{darkgreen}{HTML}{006622}
\begin{document}
\title{On the dynamical BTZ black hole in conformal gravity}
\author{Reinoud Jan Slagter\footnote{info@asfyon.com or reinoudjan@gmail.com}}
\address{Asfyon, Astronomisch Fysisch Onderzoek Nederland  \\and \\
University of Amsterdam, The Netherlands}
\begin{abstract}
We review the (2+1)-dimensional Ba\u nados-Teitelboim-Zanelli black hole solution in conformally invariant gravity, uplifted to (3+1)-dimensional spacetime. As matter content we use a scalar-gauge field.
The metric is written as $g_{\mu\nu}=\omega^2\tilde g_{\mu\nu}$, where the {\it dilaton field} $\omega$ contains all the scale dependencies and where $\tilde g_{\mu\nu}$ represents the "un-physical" spacetime.
A numerical solution is presented and shows how the dilaton can be treated on equal footing with the scalar field. The location of the apparent horizon and ergo-surface depends critically on the parameters and initial values  of the model. It is not a hard task to find suitable  initial parameters in order to obtain a regular and {\it singular free} $g_{\mu\nu}$ out of a BTZ-type solution for $\tilde g_{\mu\nu}$.

In the vacuum situation,  an {\it exact} time-dependent solution in the Eddington-Finkelstein coordinates is found, which is valid for the (2+1)-dimensional BTZ spacetime as well as for the uplifted (3+1)-dimensional BTZ spacetime.
While $\tilde g_{\mu\nu}$ resembles the standard BTZ solution with its horizons, $g_{\mu\nu}$ is {\it flat}. The dilaton field becomes an infinitesimal  renormalizable quantum field, which switches on and off Hawking radiation. This solution can be used to investigate the small distance scale of the model and the black hole complementarity issues.
It can also be used  to describe the problem how to map the quantum states of the outgoing radiation as seen by a distant observer and the ingoing by a local observer in a one-to-one way. The two observers will use a different conformal gauge.
A possible connection is made with the antipodal identification and  unitarity issues.
This research shows the power of conformally invariant gravity and can be applied to bridge the gap between general relativity and quantum field theory in the vicinity of the horizons of black holes.
\end{abstract}
\maketitle
\section{Introduction}
The last decades, several different attempts were made to unravel the spacetime topology in the vicinity of a black hole. To understand, at the tiniest scales, the black hole physics dictated by the general theory of gravity(GRT), one should incorporate  quantum mechanical effects. It is conjectured that a black hole will not live forever: it will evaporate by Hawking radiation.
This radiation can be detected by an outside observer and was originally assumed to be as thermodynamically  mixed states, because the observer has no access to the hidden sector of the black hole. The black hole could not be in a pure quantum state.
The question arises then, what will an infalling observer passing the horizon and entering the "inside" of the black hole, experience?
There must be in some way, a complementarity between the infalling and outside observer: they must map the quantum states in a one-to-one way. Are the Hawking particles entangled at different sides of the horizon (in the Penrose diagram region I and II of figure 1)?
Many attempts were made to describe  the obscure behavior of the quantum black hole.
One must, however,  keep in mind, that Nature will  work along logical rules with a minimum of assumptions. If possible, one should try to avoid, by maximally extending region I of the Penrose diagram, the introduction of firewalls, white holes, Einstein Rosen-bridges or even other universes. It is  desirable not  to enter  unaccessible domains of physics. Nice overviews are  given, for example, by Polchinski\cite{pol:2016} and Almheiri\cite{alm:2013}.
Quite recently, new boundary conditions were formulated in order to preserve locality, unitarity and loss of information\cite{thooft:2016,thooft:2018,thooft:2019}.
These modified boundary conditions can be formulated by antipodal identification in the M\"{o}bius transformation group, which are one-to-one, onto and conformal. The Hawking evaporation process can then neatly described if one allow time-inversion when one travels  from region I into region II of the Penrose diagram\footnote{ a kind of M\"{o}bius-strip  embedding}. It makes the "interior" of the black hole superfluous. In fact, one stays  on the horizon and the antipodal identification take place on the horizon, which is the stereographic projection of the 2-sphere\footnote{see section 5}.
One also avoids the problem of gravitational back reaction  if one would introduce a wormhole construction or an "other universe" when entering region II. For a local observer, this effect is invisible, while the outside observer will still experience CPT invariance. He will eventually experience this process in the far future.
One should be carefully, if  one admits the dynamical evolution of the horizon(s).
We shall see in our dynamical model,  that the future and past apparent horizons will approach each other and are antipodal.

\begin{wrapfigure}{l}{0.5\linewidth}
\centering
\begin{tikzpicture}[scale=0.4]
\node (I)    at ( 4,0)   {I};
\node (II)   at (-4,0)   {II};
\node (III)  at (0, 2.5) {III};
\node (IV)   at (0,-2.5) {IV};
\path  
  (II) +(90:4)  coordinate  (IItop)
       +(-90:4) coordinate (IIbot)
       +(0:4)   coordinate (IIright)
       +(180:4) coordinate (IIleft)
       ;
\draw  (IIleft) -- (IItop) -- (IIright) -- (IIbot) -- (IIleft) -- cycle;
\path 
   (I) +(90:4)  coordinate[label=90:$i^+$] (Itop)
       +(-90:4) coordinate[label=-90:$i^-$] (Ibot)
       +(180:4) coordinate (Iright)
       +(0:4)   coordinate[label=180:$i^0$] (Ileft)
       ;
\draw (Ileft) --
          node[midway, above left]    {$\cal{J}^+$}
     (Itop) --
      (Iright) --
      (Ibot) --
          node[midway, below left]    {$\cal{J}^-$}
      (Ileft) -- cycle;
\draw[decorate,decoration=zigzag] (IItop) -- (Itop)
      node[midway, above, inner sep=2mm] {$r=0$};
\draw[decorate,decoration=zigzag] (IIbot) -- (Ibot)
      node[midway, below, inner sep=2mm] {$r=0$};
      \node (A) at (-1,-6) {};
\end{tikzpicture}
\vspace{-1cm}
\caption{{\it Plot of maximally  extended eternal Schwarzschild black hole without Hawking radiation.}}
\end{wrapfigure}
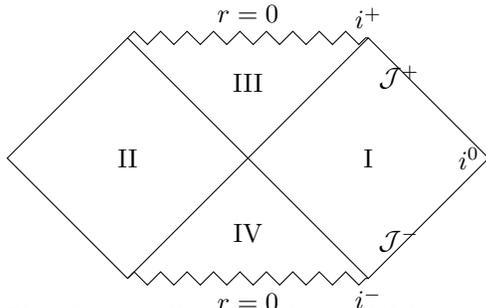

Besides the well-studied Schwarzschild and Kerr solution in general relativity theory (GRT), there is another black hole solution in $(2+1)$-dimensional spacetimes, i.e., the Ba\u nados-Teitelboim-Zanelli (BTZ) black hole\cite{banadoz:1993, carlib:1995}. The BTZ geometry solves Einstein's equations  with a negative cosmological constant in $(2+1)$-dimensions.
In general, $(2+1)$-dimensional gravity has been widely recognized as a laboratory not only for studying GRT, but also {\it quantum-gravity} models. A nice overview of these models can be found in the book of Comp\`{e}re\cite{compere:2019}.

The (2+1)-dimensional BTZ solution is comparable with the spinning point particle solution (or "cosmon"\cite{deser:1992}) of the dimensional reduced spinning cosmic string or Kerr solution.
$(2+1)$-dimensional gravity without matter, implies that the Ricci- and Riemann tensor vanish, so matter-free regions are flat pieces of spacetime. When locally a mass at rest  is present, it cuts out a wedge from the 2-dimensional space surrounding it and makes the space conical. The angle deficit is then proportional to the mass\cite{garfinkle:1985}.
The important fact is that the spinning point particle has a physical acceptable counterpart in $(3+1)$-dimensions, i.e., the spinning cosmic string. The z-coordinate is suppressed, because there is no structure
in that direction altogether. It is not a surprise that these models are used in constructing quantum gravity models. In these models one uses locally Minkowski spacetime, so {\it planar gravity} fits in very well.

The BTZ solution is related to  the Anti-deSitter/Conformal Field Theory (AdS/CFT) correspondence\cite{mald:1998} and became a tool to understand black hole entropy\cite{strom:1997}.
For the (2+1)-dimensional BTZ black hole solution, one can try to follow the same procedure as used for the cosmic string, by {\it uplifting} the solution to
(3+1)-dimensional spacetime. However, the cosmological constant must be taken zero, when the BTZ solution is uplifted, so it loses its connection with the asymptotic $AdS_3$ black hole. This opens the way to new solutions, which was done in a {\it conformally invariant} setting\cite{slagter:2019,slagter:2020}.
Conformal invariance (CI) was originally introduced by Weyl\cite{weyl:1918}. See also the text book of Wald\cite{wald:1984}.
The AdS/CFT   correspondence renewed the interest in conformal gravity. AdS/CFT is a conjectured relationship between two kinds of physical theories. AdS spaces  are used in theories of quantum gravity while  CFT  includes theories similar to the Yang Mills theories that describe elementary particles.
It is believed that CI can help us to move a little further along the road to quantum gravity.
Exact local CI at the level of the Lagrangian, will then spontaneously be broken, comparable with the Brout-Englert-Higgs (BEH) mechanism.
It is an approved alternative for disclosing the small-distance structure when one tries to describe quantum-gravity problems\cite{thooft:2010,thooft:2011,thooft:2011b}.  It  can also be used to model scale-invariance in the  cosmic microwave background radiation (CMBR)\cite{bars:2014}. Another interesting application  can be found in the work of Mannheim on conformal cosmology\cite{mannheim:2005}. This model could serve as an alternative approach to explain  the rotational curves of galaxies, without recourse to dark matter and dark energy (or cosmological constant).
The key problem is the handling of asymptotic flatness of isolated systems in GRT, specially when they radiate and the generation of the metric $g_{\mu\nu}$ from at least Ricci-flat spacetimes.
In the non-vacuum case one should construct a Lagrangian where spacetime and the fields defined on it, are topological regular and physical acceptable. This can be done by considering the scale factor (or {\it warp factor} in higher-dimensional models\cite{slagter:2016}) as a {\it dilaton field} besides, for example, a conformally coupled  scalar field or other fields.
Conformal invariant gravity distinguishes itself by the notion that the spacetime is written as $g_{\mu\nu}=\omega^2 \tilde g_{\mu\nu}$, with $\omega$ a dilaton field which contains all the scale dependencies  and $\tilde g_{\mu\nu}$ the {\it "un-physical"} spacetime, related to the $(2+1)$-dimensional Kerr and BTZ black hole solution.
$\omega$ is just an ordinary renormalizable field, which  could create the spacetime twofold: an infalling and outside observer  use different ways to fix the conformal gauge in order to overcome the {\it unitarity problems} encountered in standard approaches in quantum gravity models. It can be handled on equal footing with a scalar field.
Renormalization and unitarity problems in general relativity at the quantum scale, have a long history\cite{veltman:1974,stelle:1977}.
In first instance, it was believed that  conformally invariance would not survive in quantum gravity (see, for example, the overview of Duff\cite{duff:1993}). However, new interest occurred, when it was realized that  Weyl anomalies and unitarity problems could be overcome.
In constructing an effective theory in canonical quantum gravity and to obtain quantum amplitudes, one performs a functional integration of the exponent of the entire action over, for example, all components of the metric tensor at all spacetime points.
Now the integration is first performed over the dilaton function $\omega$ together with the matter fields\cite{thooft:2011b}. Integration over the $\omega$ is identical to the integration over a renormalizable scalar field. In the action the dilaton must be  shifted to the complex contour, in order to obtain the same unitarity and positivity features as the scalar field.

As already mentioned, another actual problem is the back hole {\it complementarity}: how to handle the in- and out-going radiation as experienced by an infalling-  and  outside observer.
For the infalling observer, it takes a finite proper time to cross the horizon, while an outside observer never sees them on the horizon for an infinite time.
In a dynamical setting, there will be a back-reaction on the location of the horizon(s). The in falling and outside observer will experience a different $\omega$. They use different ways to fix the conformal gauge.
It is conjectured that the issues of unitarity, locality and the antipodal identification are related to the dilaton field $\omega$ in our conformal invariant model using the conformal compactification.

In section 2 and 3  we describe the dynamical CI model on the original BTZ black hole spacetime, uplifted to (3+1) dimensions. We also present  a numerical solution of the complete set of coupled PDE's.
In section 4 and 5 we find an exact time dependent solution in the vacuum situation in Eddington-Finkelstein coordinates and we explain possible ways to connect this solution with recent research on black hole complementarity, antipodal identifications and hawking radiation. In section 6 we explain briefly the M\"{o}bius group.
\section{The BTZ solution revised}
If one solves the Einstein equations $G_{\mu\nu}=\lambda g_{\mu\nu}$ for the spacetime
\begin{eqnarray}
ds^2=-N(\rho)^2 dt^2+\frac{1}{N(\rho)^2}d\rho^2+\rho^2\Bigl(d\varphi+N^\varphi(\rho)dt\Bigr)^2,\label{2-1}
\end{eqnarray}
one obtains
\begin{eqnarray}
N(\rho)^2\equiv \alpha^2-\Lambda \rho^2+\frac{16 G^2J^2}{\rho^2}  , \qquad  N^\varphi(\rho)\equiv -\frac{4GJ}{\rho^2} +S\label{2-2},
\end{eqnarray}
where $S$, $J$ and $\alpha$ are integration constants\cite{banadoz:1993,compere:2019}.  The parameters $\alpha$ and $J$  represent the standard ADM mass ($\alpha^2 =\pm 8GM$) and angular momentum and determine the asymptotic behavior of the solution. $\Lambda$ represents the cosmological constant. There is an inner and outer horizon  and an ergo-circle just as in the case of the Kerr spacetime. However, if one lifts-up this spacetime to $(3+1)$ dimensions, one must take $\Lambda=0$, which can easily be verified by the Einstein equations. So we consider here the case $\Lambda=0$, and we write the spacetime as
\begin{eqnarray}
ds^2=-\Bigl[8G(JS-M)-S^2\rho^2\Bigr]dt^2+\frac{\rho^2 r_H^2}{16G^2J^2(\rho_H^2-r^2)}d\rho^2+\rho^2 d\varphi^2 \cr
\qquad\qquad\qquad+2\rho^2\Bigl(S-\frac{4GJ}{\rho^2}\Bigr)dt d\varphi\label{2-3},
\end{eqnarray}
with $\rho_H$ the horizon $\rho_H=\sqrt{\frac{2G}{M}}J$.
In the case of $S=0$, which is also done  in the original BTZ solution, one can transform the spacetime to
\begin{equation}
ds^2=-\Bigl(\alpha dt+\frac{4GJ}{\alpha}d\varphi \Bigr)^2+d(\rho')^2+\alpha^2 (\rho')^2 d\varphi^2, \label{2-4}
\end{equation}
by $(\rho')^2=\frac{16G^2J^2+\alpha^2 \rho^2}{\alpha^4}$.
This is just the spinning particle spacetime \cite{deser:1992}.

In a former study\cite{slagter:2019}, we investigated the revised BTZ solution in connection with the spinning cosmic strings and conformally invariance and found an uplifted exact vacuum solution.
Here we extend this study.
\section{The dynamical BTZ  model}
\subsection{The field equations}
Let us consider the time-dependent spacetime $g_{\mu\nu}\equiv \omega(t,\rho)^2\tilde g_{\mu\nu}$
\begin{equation}
ds^2=\omega(t,\rho)^2\Bigl[-N(t,\rho)^2 dt^2+\frac{1}{N(t,\rho)^2}d\rho^2+dz^2+\rho^2\Bigl(d\varphi+N^\varphi(t,\rho)dt\Bigr)^2\Bigr]\label{3-1},
\end{equation}
with $\omega$ the dilaton field.
The action under consideration is
\begin{eqnarray}
{\cal S}=\int d^4x\sqrt{- \tilde g}\Bigl\{-\frac{1}{12}\Bigl(\tilde\Phi\tilde\Phi^*+\bar\omega^2\Bigr) \tilde R-\frac{1}{2}\Bigl( D_\alpha\tilde\Phi( D^\alpha\tilde\Phi)^*+\partial_\alpha\bar\omega\partial^\alpha\bar\omega\Bigr)\cr
-\frac{1}{4}F_{\alpha\beta}F^{\alpha\beta}-V(\tilde\Phi ,\bar\omega)-\frac{1}{36}\kappa^2\Lambda\bar\omega^4\Bigr\}\label{3-2}.
\end{eqnarray}
We parameterize the scalar and gauge field as
\begin{eqnarray}
A_\mu=\Bigr[P_0(t,\rho),0,0,\frac{1}{e}(P(t,\rho)-n)\Bigr], \qquad \tilde\Phi(t,\rho)=\eta X(t,\rho)e^{in\varphi}\label{3-3}.
\end{eqnarray}
The gauge covariant derivative is $D_\mu\Phi=\tilde \nabla_\mu\Phi+ie A_\mu\Phi$ and $F_{\mu\nu}$ the abelian field strength.

In the action one redefined $\bar\omega^2 \equiv-\frac{6\omega^2}{\kappa^2}$ (in order to ensure that the $\omega$ field has the same unitarity and positivity properties as the scalar field $\Phi$\cite{thooft:2015})  and $\Phi=\frac{1}{\omega}\tilde\Phi$.
This Lagrangian  is local conformally invariant under the transformation $\tilde g_{\mu\nu}\rightarrow\Omega^2 \tilde g_{\mu\nu}, \tilde \Phi \rightarrow \frac{1}{\Omega}\tilde \Phi$ and $\bar\omega\rightarrow \frac{1}{\Omega}\bar\omega$.

Varying the Lagrangian with respect to $\tilde g_{\mu\nu}, \tilde \Phi, \bar\omega$ and $A_\mu$, we obtain the equations
\begin{eqnarray}
\tilde G_{\mu\nu}=\frac{1}{(\bar\omega^2 +\tilde\Phi\tilde\Phi^*)}\Bigl(\tilde T_{\mu\nu}^{(\bar\omega)}+\tilde T_{\mu\nu}^{(\tilde\Phi,c)}+\tilde T_{\mu\nu}^{(A)}+\frac{1}{6}\tilde g_{\mu\nu}\Lambda\kappa^2\bar\omega^4
+\tilde g_{\mu\nu}V(\tilde\Phi,\bar\omega)\Bigr),\label{3-4}
\end{eqnarray}
\begin{eqnarray}
\tilde\nabla^\alpha \partial_\alpha\bar\omega -\frac{1}{6}\tilde R\bar\omega -\frac{\partial V}{\partial \bar\omega}-\frac{1}{9}\Lambda \kappa^2\bar\omega^3=0, \label{3-5}
\end{eqnarray}
\begin{eqnarray}
 \tilde D^\alpha \tilde D_\alpha\tilde\Phi-\frac{1}{6}\tilde R\tilde\Phi-\frac{\partial V}{\partial\tilde\Phi^*}=0,\qquad \tilde\nabla^\nu F_{\mu\nu}=\frac{i}{2}\epsilon \Bigl(\tilde\Phi ( \tilde D_\mu\tilde\Phi)^* -\tilde\Phi^*  \tilde D_\mu\tilde\Phi\Bigr),\label{3-6}
\end{eqnarray}
with
\begin{eqnarray}
\hspace{-0.5cm}
\tilde T_{\mu\nu}^{(A)}=F_{\mu\alpha}F_\nu^\alpha-\frac{1}{4}\tilde g_{\mu\nu}F_{\alpha\beta}F^{\alpha\beta},\label{3-7}
\end{eqnarray}
\begin{eqnarray}
\tilde T_{\mu\nu}^{(\tilde\Phi ,c)}=\Bigl(\tilde\nabla_\mu\partial_\nu\tilde\Phi\tilde\Phi^*-\tilde g_{\mu\nu}\tilde\nabla_\alpha\partial^\alpha\tilde\Phi\tilde\Phi^*\Bigr)\cr
-3\Bigl[ \tilde D_\mu\tilde\Phi( \tilde D_\nu\tilde\Phi)^*+( \tilde D_\mu\tilde\Phi)^* \tilde D_\nu\tilde\Phi
-\tilde g_{\mu\nu} \tilde D_\alpha\tilde\Phi( \tilde D^\alpha\tilde\Phi)^*\Bigl]\label{3-8}
\end{eqnarray}
and
\begin{eqnarray}
\hspace{-0.5cm}
\tilde T_{\mu\nu}^{(\bar\omega)}=\Bigl(\tilde\nabla_\mu\partial_\nu\bar\omega^2-\tilde g_{\mu\nu}\tilde\nabla_\alpha\partial^\alpha\bar\omega^2\Bigr)
-6\Bigl(\partial_\mu\bar \omega\partial_\nu\bar \omega-\frac{1}{2}\tilde g_{\mu\nu}\partial_\alpha\bar \omega\partial^\alpha\bar\omega)\Bigl).\label{3-9}
\end{eqnarray}
The covariant derivatives are taken with respect to $\tilde g_{\mu\nu}$.
Newton's constant reappears in the quadratic interaction term for the scalar field. One refers to the field $\bar\omega(t,\rho)$ as a dilaton field. A massive term in $V(\tilde\Phi,\bar\omega)$ will break the {\it tracelessness} of the energy momentum tensor, a necessity for conformally invariance.
The cosmological constant $\Lambda$ could be ignored from the point of view of naturalness in order to avoid the inconceivable  fine-tuning. Putting $\Lambda$ zero increases the symmetry of the model.

Note that we cannot use in the stationary CI invariant model the {\it gauge} $A_t=0$. In standard gauged vortices models, this gauge simplifies the well-known Nielsen-Olesen $n=1$ vortex solution. The spatial rotational symmetry can then be completely compensated by a spatially uniform gauge transformation. In the stationary situation this is not the case.

For the Maxwell field $A_\mu$, the equation for $A_t$ is a constraint equation and in a time-dependent  setting, only $A_i$ are dynamical. Standard, one uses then the Lorentz-gauge to remove $A_t$ completely.
Gauge invariance is necessary in order to overcome breaking of locality and unitarity.
In models with arbitrary   vorticity n and SU(2)-Yang-Mills-Higgs theory, Gauss's law yields also a non-zero $A_t$ for most gauges. Just as in the monopole and dyon solutions, $A_t$ produces a back reaction on $A_i$ perturbatively. Although the dyon fields are time-independent, there is a net kinetic energy because $A_t$ is non-vanishing, so are steadily rotating (see for example the text book of Weinberg\cite{weinberg:2012}).

In our model we have rotation, i.e., a term $N^\varphi(t,\rho)$. If we calculate the conservation equations for the Einstein equations, one easily finds that
\begin{equation}
P_0=\frac{1}{e}P N^\varphi.\label{3-10}
\end{equation}
So we obtained a kind of  natural "gauge" in order to get rid of $A_0$ (Eq.(\ref{3-3})). The equation for $N^\varphi$ decouples from the other equations.

The field equations now become
\begin{eqnarray}
\ddot N=-N^4(N''+3\frac{N'}{\rho})+3\frac{\dot N^2}{N}-N^3N'^2+\frac{1}{\eta^2 X^2+\omega^2}\Bigl[3N(N^4\omega'^2-\dot\omega^2)\cr
-N^3V-\frac{N^5}{\rho}(\omega\omega'+\eta^2XX')+3\eta^2N(N^4X'^2-\dot X^2)+2\frac{N}{e^2\rho^2}(\dot P^2-N^4 P'^2)\cr
-6\frac{N^4}{e^2\rho^3}N'-\frac{1}{6}\kappa^2\Lambda N^3\omega^4+6\frac{\eta^2P^2N^3X^2}{\rho^2}\Bigr]+\frac{1}{(\eta^2 X^2+\omega^2)^2}\Bigl[18\frac{\eta^2P^4X^2N^3}{e^2\rho^4}\cr
+3\frac{NP^2}{e^4\rho^4}(\dot P^2-N^4 P'^2)-6\frac{P^2N^5}{e^2\rho^3}(\omega\omega'+\eta^2 XX')\Bigr],\qquad\qquad\label{3-11}
\end{eqnarray}
\begin{eqnarray}
\ddot \omega=N^4(\omega''+\frac{\omega'}{\rho})+2N^3\omega'N'+2\frac{\dot \omega\dot N}{N}+\frac{\eta XN^2}{\omega}\frac{dV}{dX}  \qquad\qquad\qquad\cr
+\frac{\Bigl[\omega(N^4 \omega'^2-\dot \omega^2)+\eta^2\omega(N^4 X'^2-\dot X^2)-2\frac{\omega P^2N^3}{e^2\rho^3}N'\Bigr]}{\eta^2X^2+\omega^2} \qquad\qquad \cr
+\frac{1}{(\eta^2X^2+\omega^2)^2}\Bigl[-2\frac{\omega P^2N^4}{e^2\rho^3}(\omega\omega'+\eta^2 XX')
+\frac{1}{2e^2r^2}\omega (\dot P^2-N^4 P'^2)\cr
\cdot(\omega^2+2\frac{P^2}{e^2\rho^2}+\eta^2X^2)
+\frac{\eta^2 P^2N^2X^2\omega}{\rho^2}(\omega^2+6\frac{P^2}{e^2\rho^2}+\eta^2X^2)\cr
-\frac{N^2}{3\omega}(V+\frac{1}{6}\kappa^2\Lambda\omega^4)(3\omega^4+5\eta^2X^2\omega^2+2\eta^4X^4)\Bigr],\qquad\label{3-12}
\end{eqnarray}
\begin{eqnarray}
\ddot X=N^4(X''+\frac{X'}{\rho})+2N^3X'N'+2\frac{\dot X\dot N}{N}-\frac{N^2}{\eta}\frac{dV}{dX}  \qquad\qquad\cr
+\frac{\Bigl[\eta^2 X(N^4 X'^2-\dot X^2)+X(N^4\omega'^2-\dot\omega^2)-2\frac{XP^2N^3}{e^2\rho^3}N'
-\frac{XN^2}{3}(V+\frac{1}{6}\kappa^2\Lambda \omega^4\Bigr]}{\eta^2X^2+\omega^2} \cr
+\frac{1}{(\eta^2X^2+\omega^2)^2}\Bigl[-2\frac{XP^2N^4}{e^2\rho^3}(\omega\omega'+\eta^2 XX')
+\frac{1}{2e^2\rho^2}X(\dot P^2-N^4 P'^2)\cr
\cdot(\omega^2+2\frac{P^2}{e^2\rho^2}+\eta^2X^2)
-\frac{P^2N^2X}{\rho^2}(\omega^4-6\frac{\eta^2X^2P^2}{e^2\rho^2}+\eta^2X^2\omega^2)\Bigr],\qquad\qquad\label{3-13}
\end{eqnarray}
\begin{eqnarray}
\ddot P=N^4(P''-\frac{P'}{\rho})+e^2\eta^2PX^2N^2+N^3P'N'+2\frac{\dot P\dot N}{N}-4\frac{PN^3N'}{r} \cr
+2\frac{P \Bigl[ \dot P^2-N^4P'^2-2\rho e^2N^4(\eta^2 XX'+\omega \omega')+6\eta^2 e^2N^2X^2P^2\Bigr] }{\rho^2e^2(\eta^2 X^2+\omega^2)},\label{3-14}
\end{eqnarray}

\begin{eqnarray}
(\dot N^\varphi)'=-(N^\varphi)'\frac{\dot P}{P},\qquad  (N^\varphi)''=-(N^\varphi)'(\frac{P'}{P}+\frac{1}{\rho}),\label{3-15}
\end{eqnarray}
\begin{equation}
{(N^\varphi)'}^2=-4\frac{N}{\rho^3}N'+2\frac{\Bigl[\frac{\dot P^2}{N^2}-N^2P'^2+6e^2\eta^2X^2P^2-2e^2 \rho N^2(\eta^2XX'+\omega\omega')\Bigr]   }{e^2\rho^4(\eta^2X^2+\omega^2)}.\label{3-16}
\end{equation}
Further, we obtain from the Maxwell equations and Einstein constraint equations
\begin{eqnarray}
\dot P=\frac{2\eta^2X\dot X+\omega\dot\omega}{\eta^2X^2+\omega^2-2\frac{P^2}{e^2\rho^2}},\quad
P'=\frac{2\eta^2X X'+\omega\omega'}{\eta^2X^2+\omega^2-2\frac{P^2}{e^2\rho^2}}.\label{3-17}
\end{eqnarray}
The equation for $\omega$ is obtained from the Einstein equations and the scalar equation for X. If we substitute back the equations into the dilaton equation, we obtain the relation for the potential
\begin{equation}
\frac{2}{3}V=\eta X\frac{dV}{d X}+\omega\frac{dV}{d\omega},\label{3-18}
\end{equation}
From the conservation equations we then obtain
\begin{eqnarray}
\dot V=5\frac{\eta^2X^2P\dot P}{\rho^2}+6\eta\dot X\frac{dV}{dX}+6\dot\omega\frac{dV}{d\omega}, \cr
V'=5\frac{\eta^2X^2P P'}{\rho^2}+6\eta X'\frac{dV}{dX}+6\omega'\frac{dV}{d\omega}.\label{3-19}
\end{eqnarray}
Sometimes,   one chooses a unitary gauge in order to obtain a comparable relation  (see for example Oda\cite{oda:2015})
\subsection{The numerical solution}
We can plot a numerical solution of the field equations of section 3.1 for a set of initial and boundary values. We can choose as initial values the {\it vacuum solution} of Eq.(\ref{3-11})-(\ref{3-16}). This solution can easily be found exactly:
\begin{figure}[h]
\centerline{
\includegraphics[width=4.3cm]{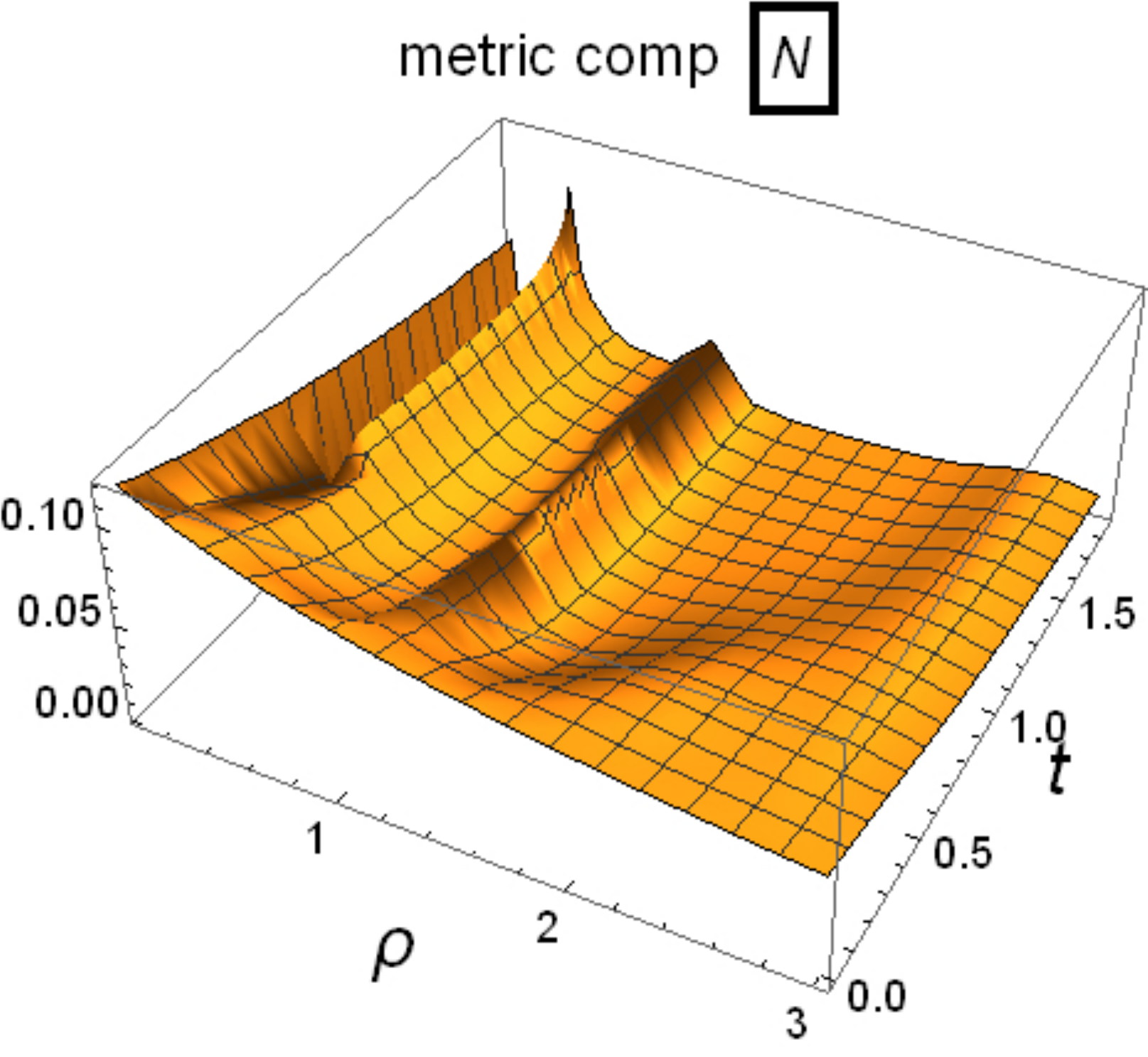}
\includegraphics[width=4.3cm]{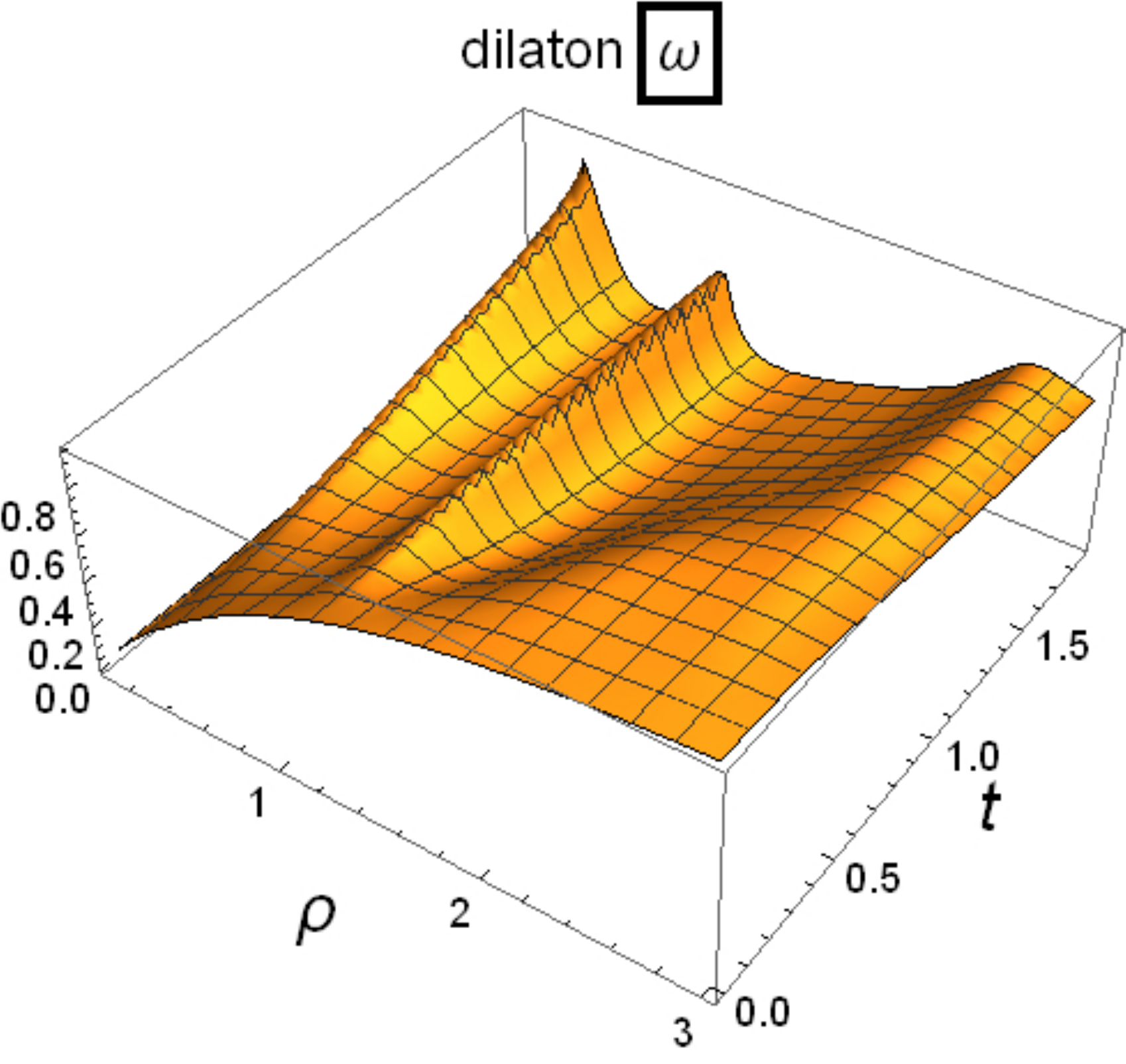}
\includegraphics[width=4.3cm]{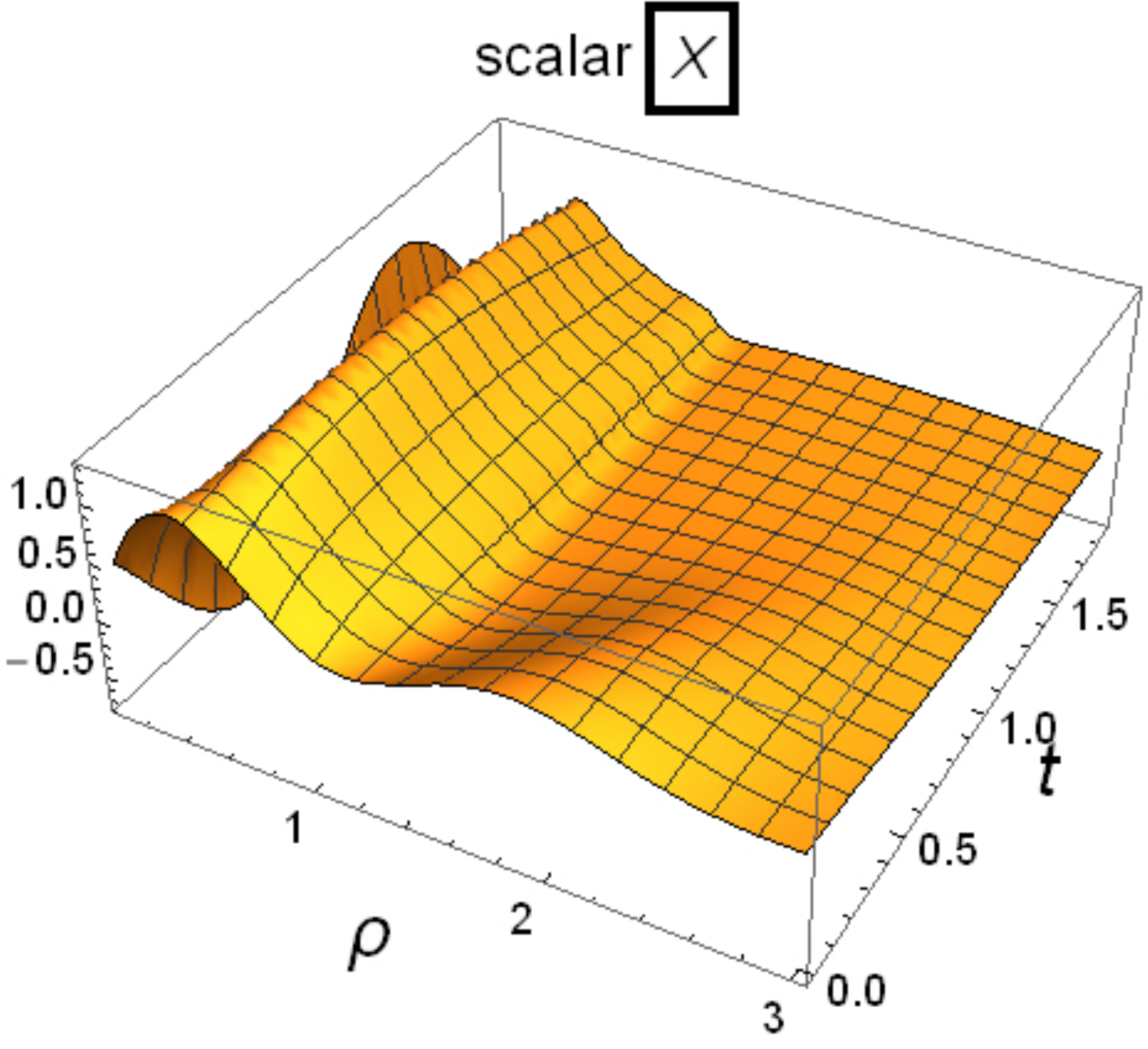}}
\centerline{
\includegraphics[width=4.3cm]{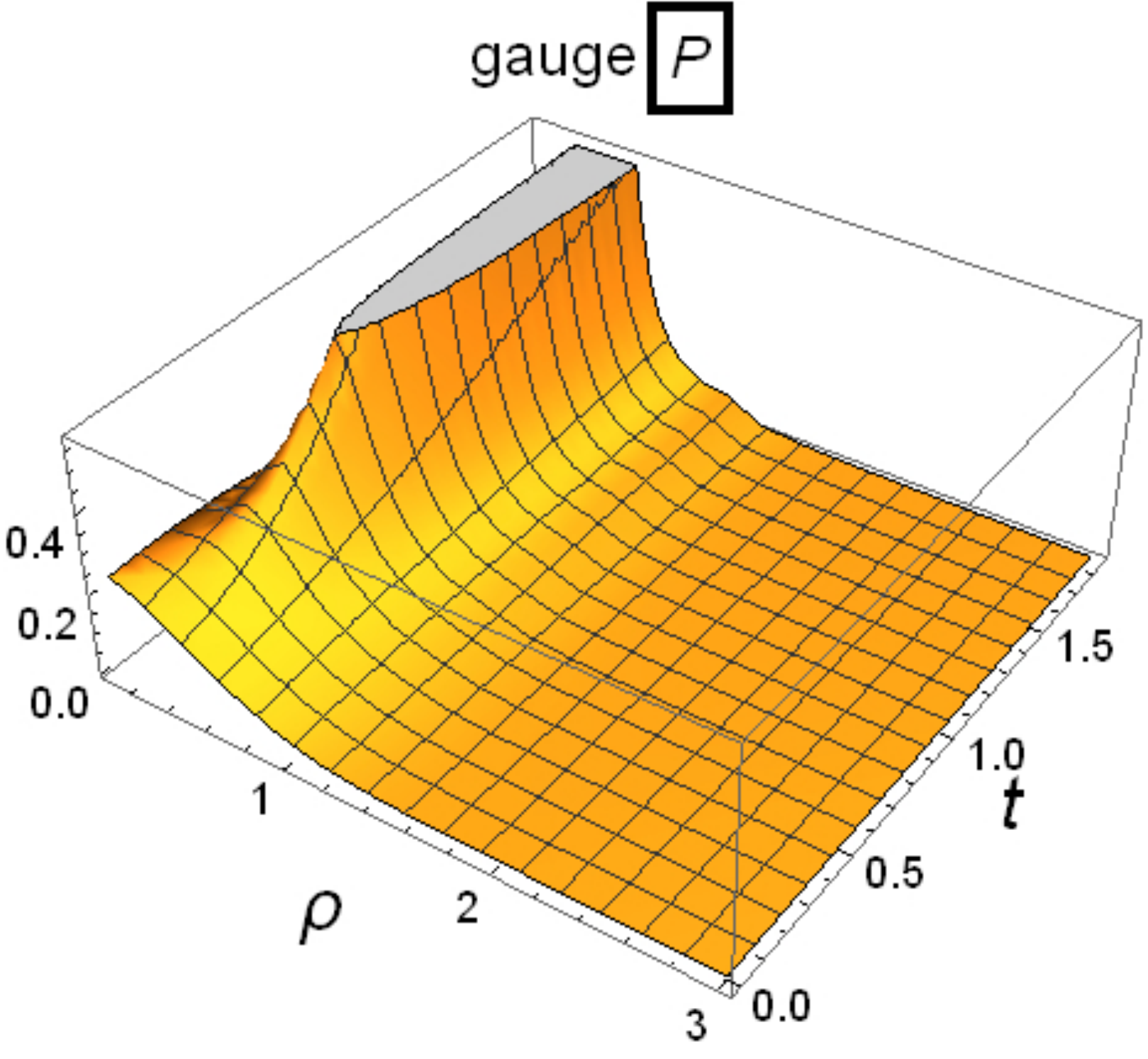}
\includegraphics[width=4.3cm]{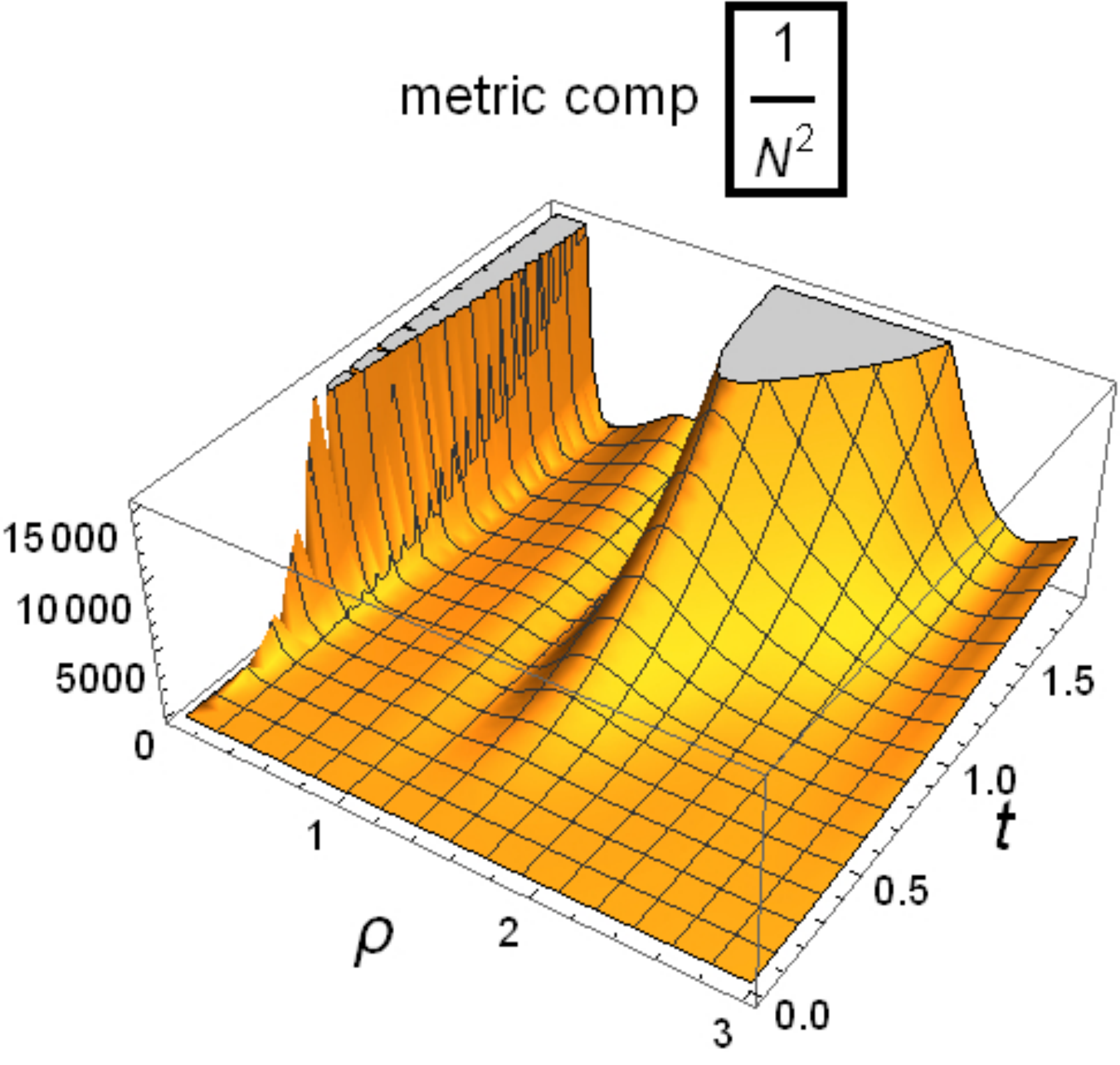}
\includegraphics[width=4.3cm]{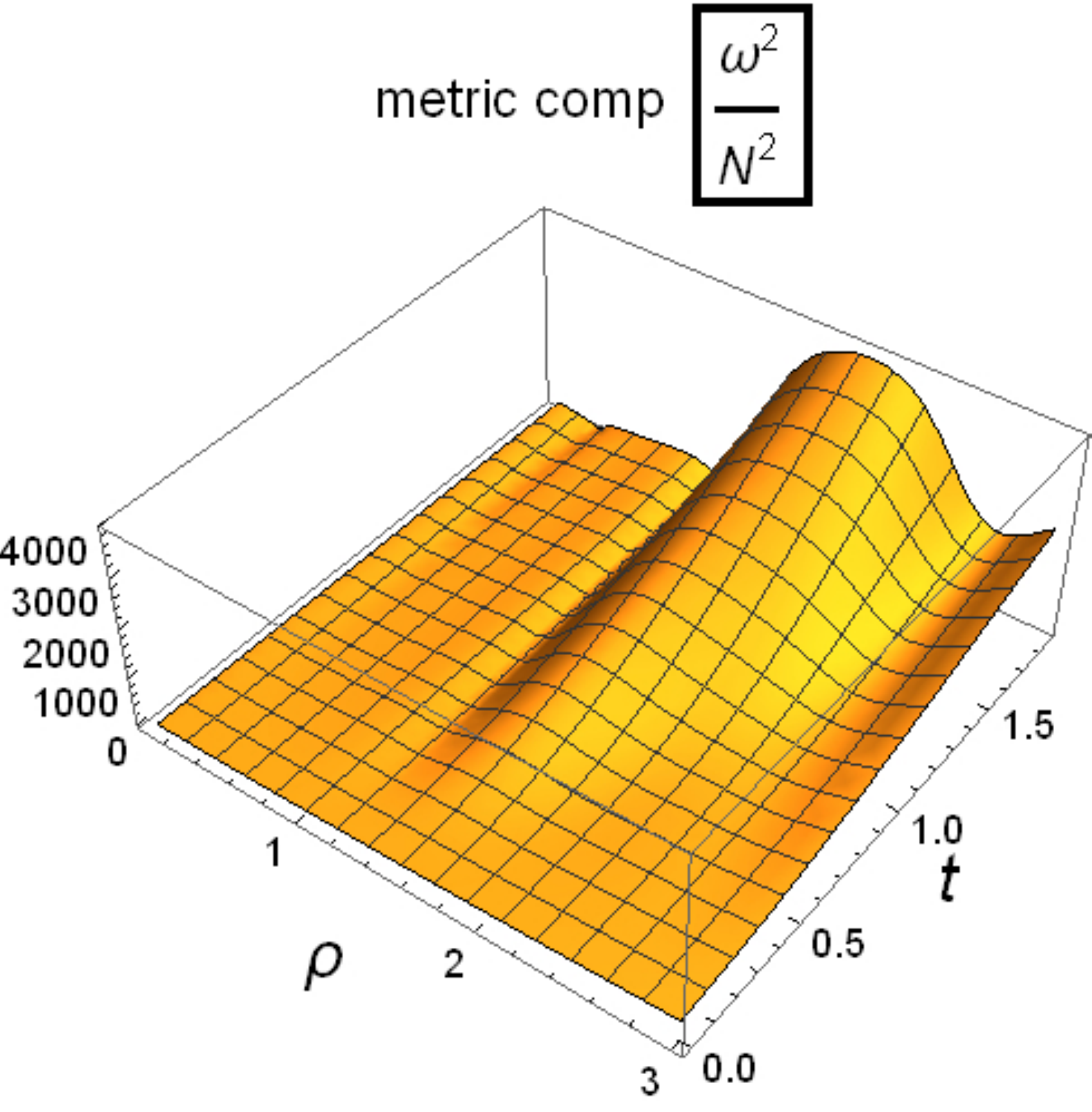}}
\centerline{
\includegraphics[width=4.3cm]{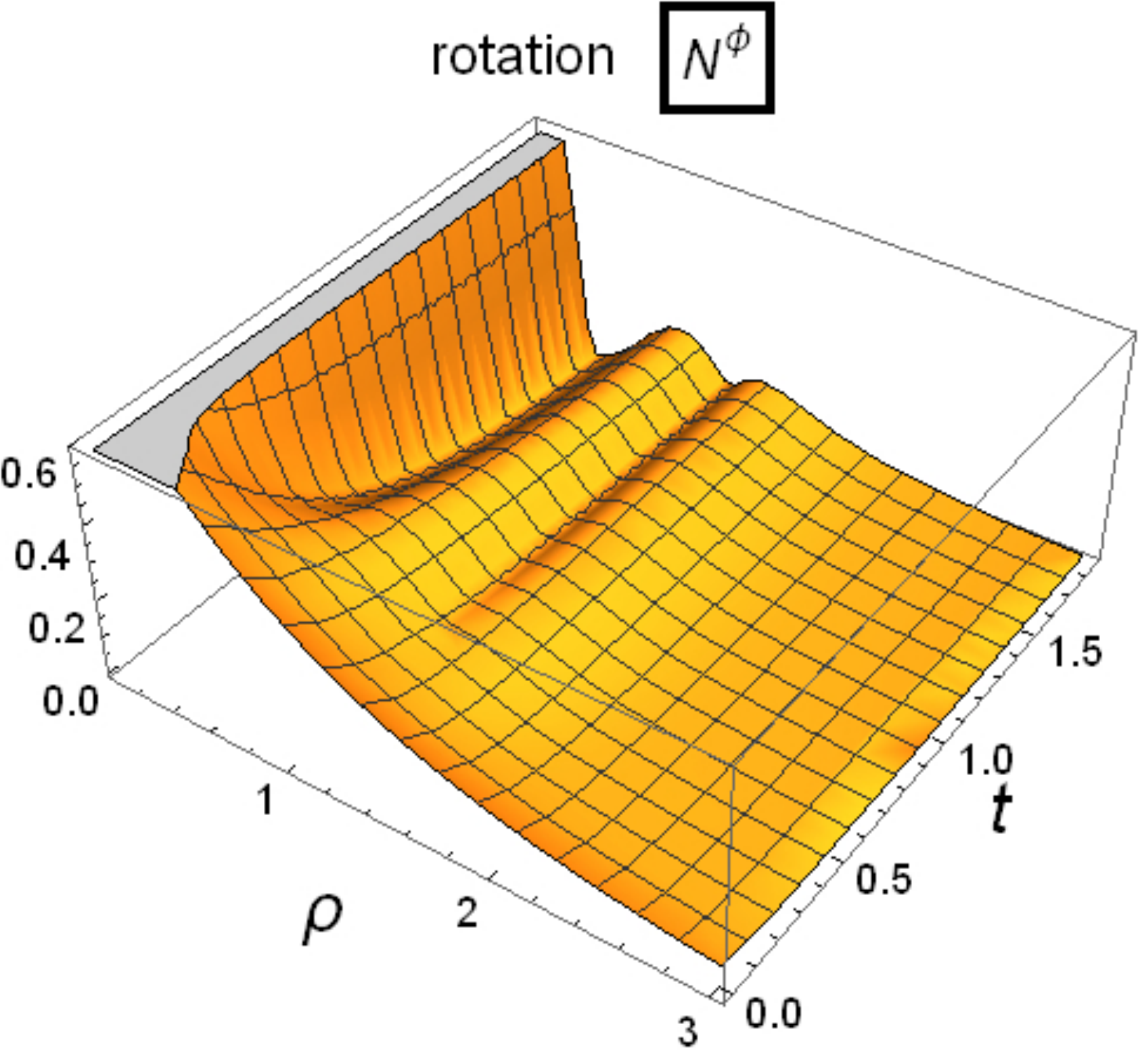}
\includegraphics[width=4.3cm]{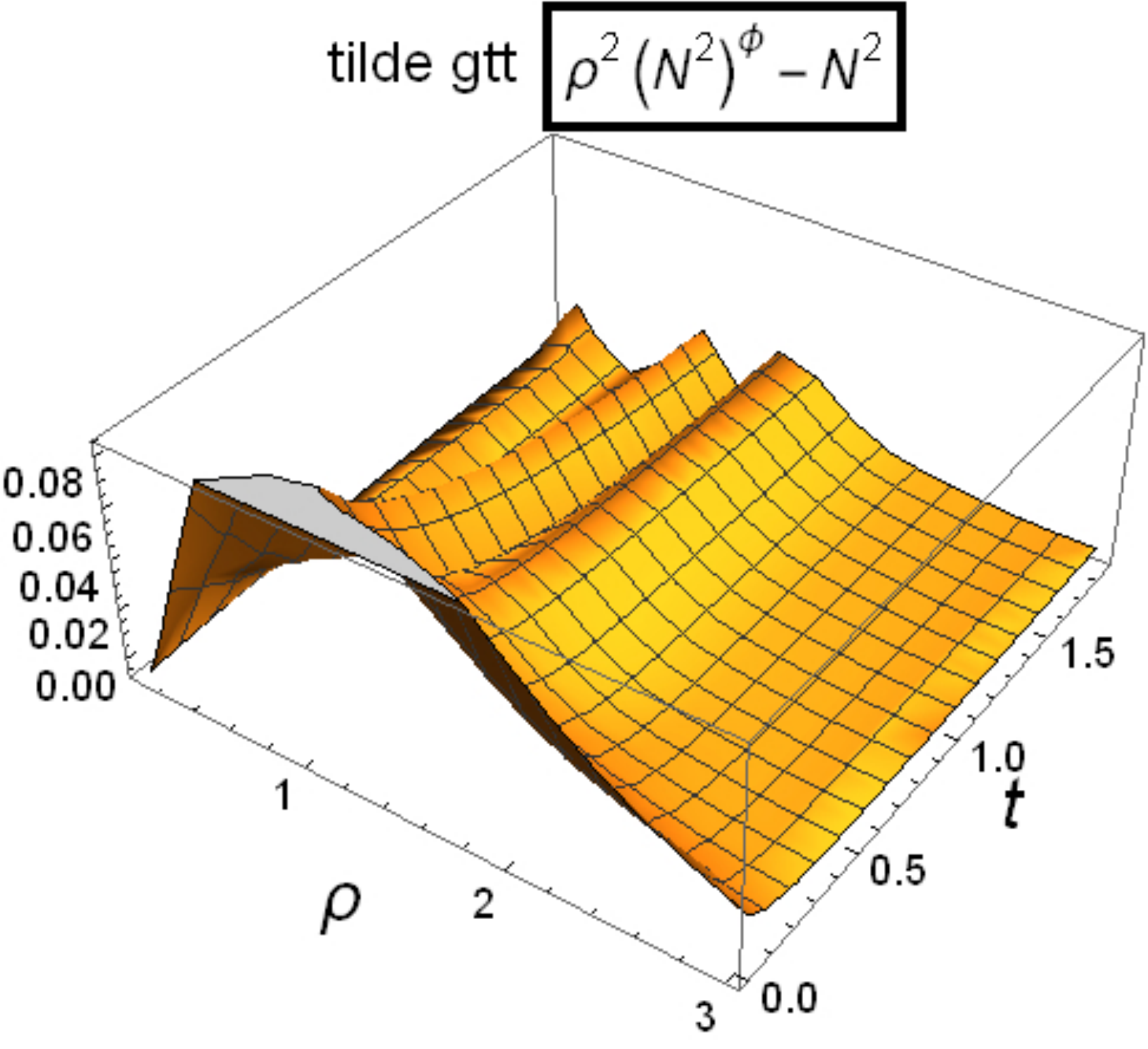}
\includegraphics[width=4.3cm]{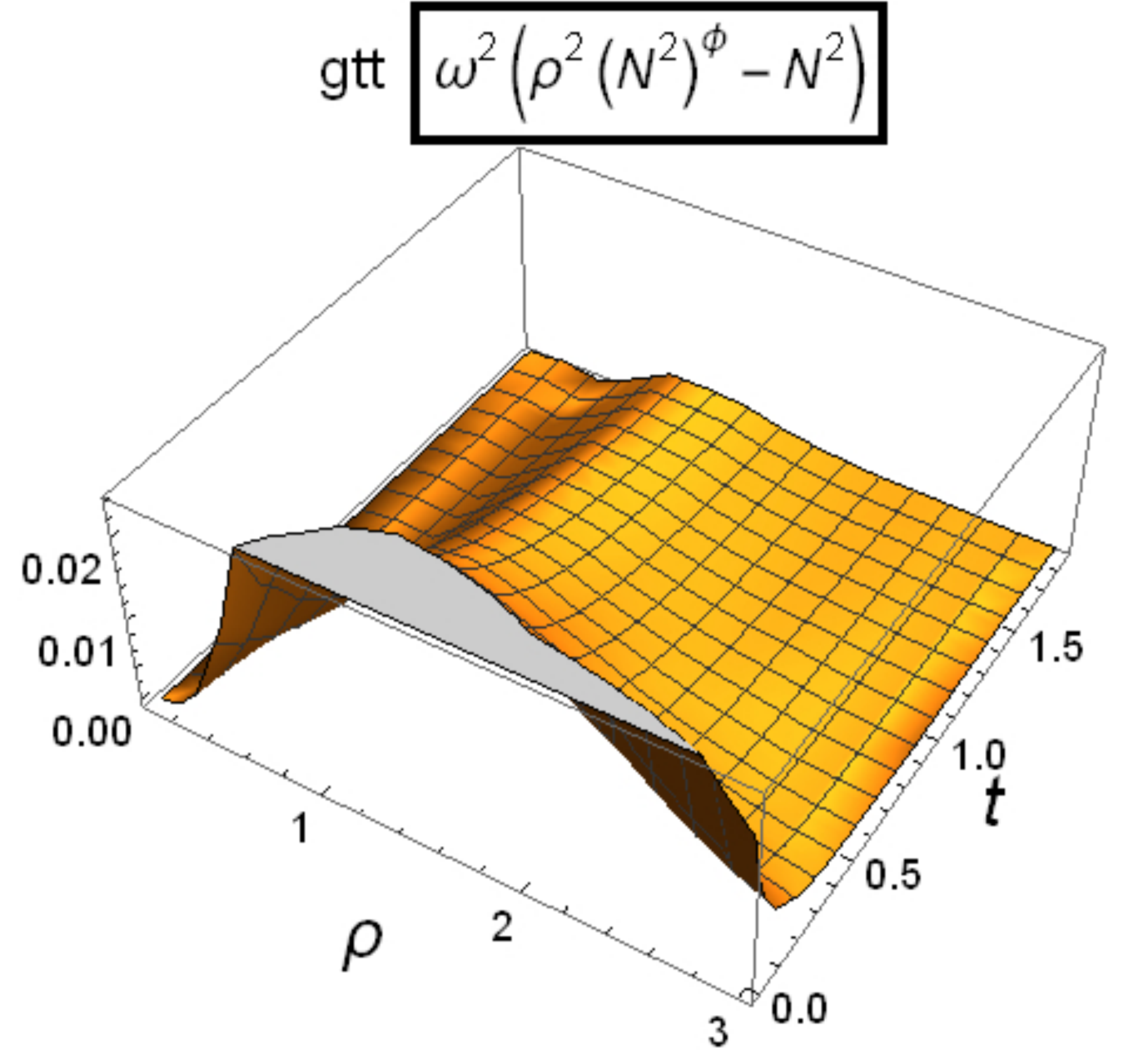}}
\vspace{1.cm}
\caption{{\it Example of a numerical solution of the system of Eq.(\ref{3-11})-(\ref{3-16}) with only for the scalar field X an outgoing wavelike initial value and a increasing mode for the dilaton. We used the potential from Eq.(\ref{3-18}).}  }
\end{figure}
\begin{figure}[h]
\centerline{
\includegraphics[width=4.3cm]{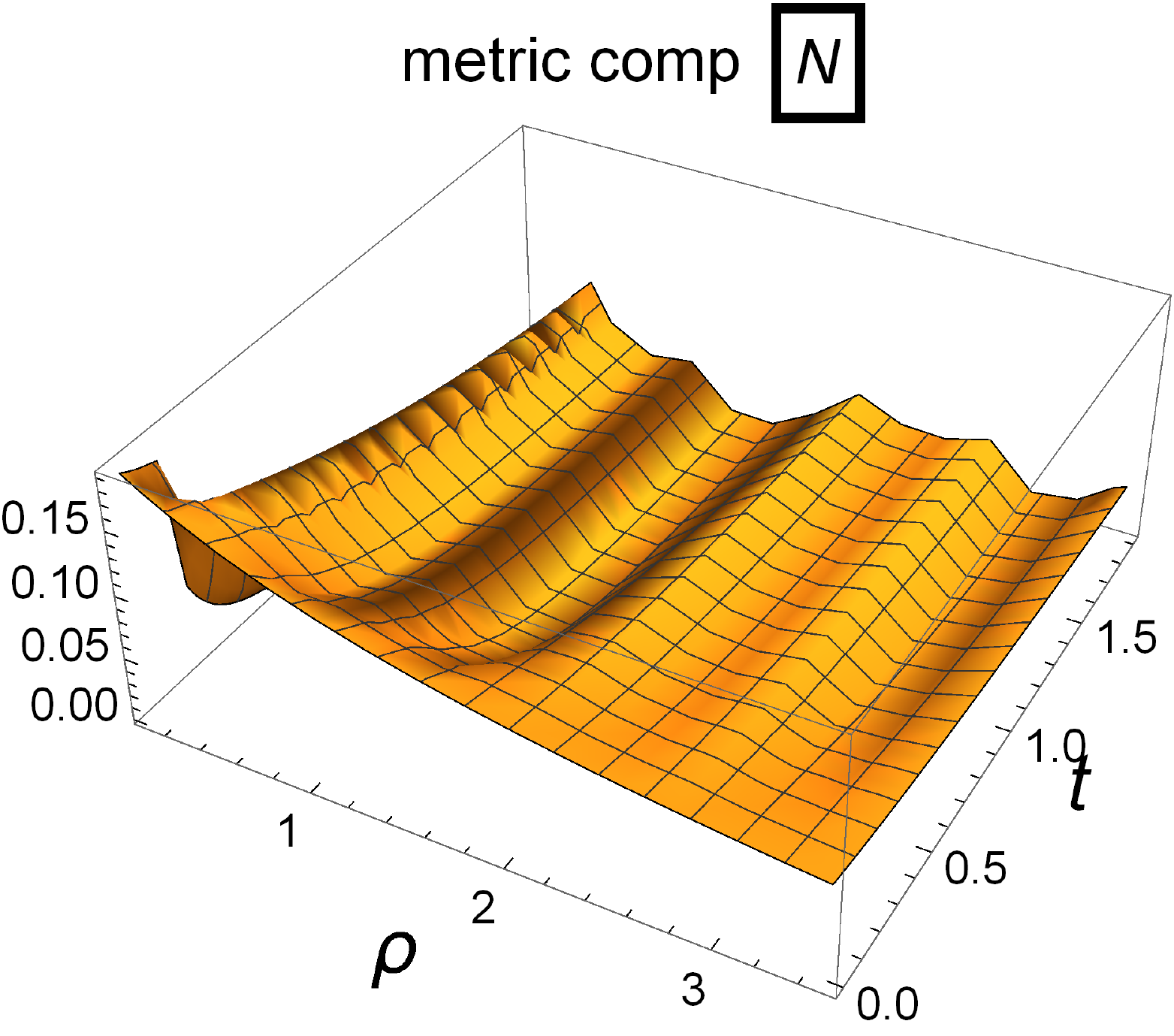}
\includegraphics[width=4.3cm]{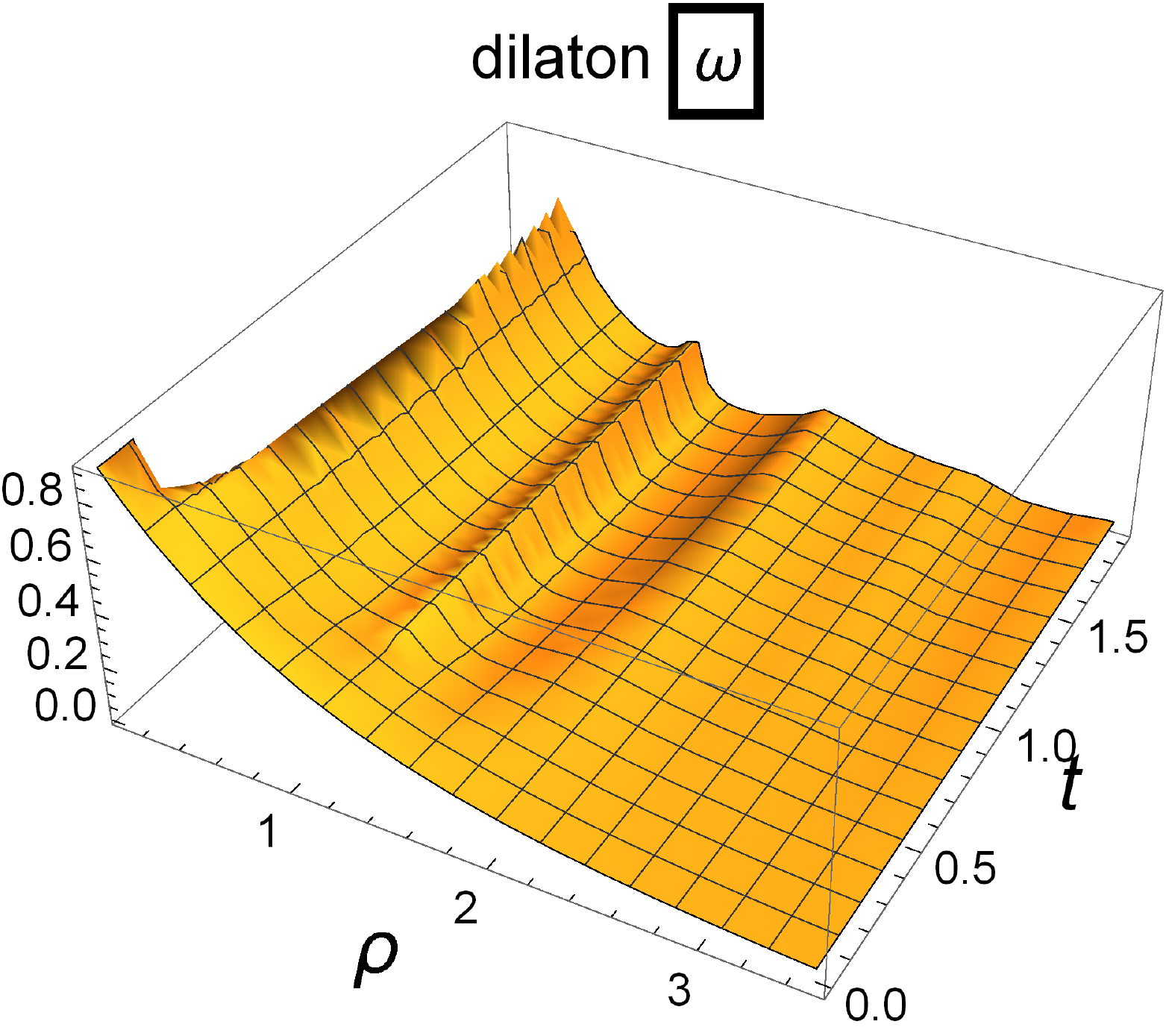}
\includegraphics[width=4.3cm]{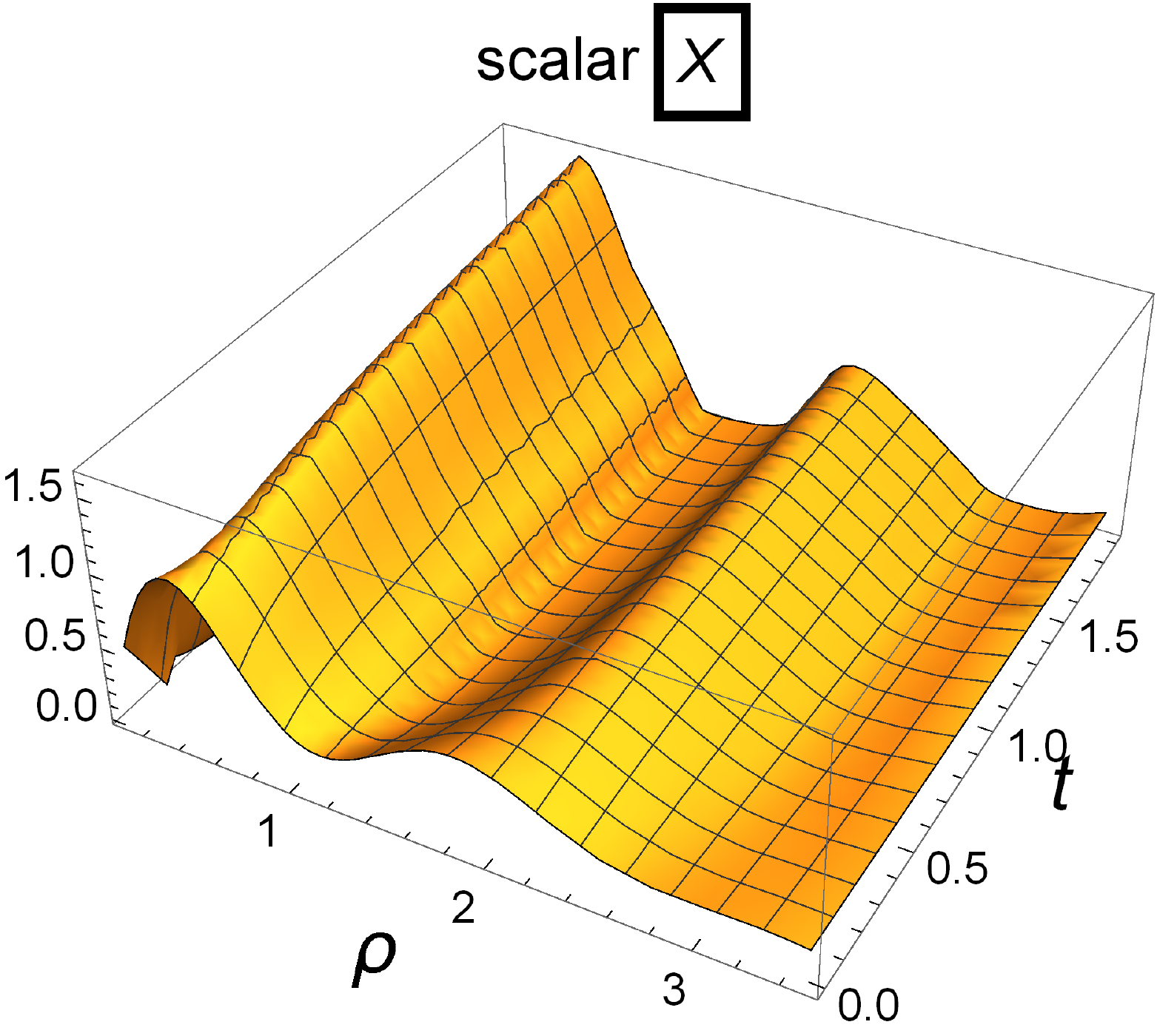}}
\centerline{
\includegraphics[width=4.3cm]{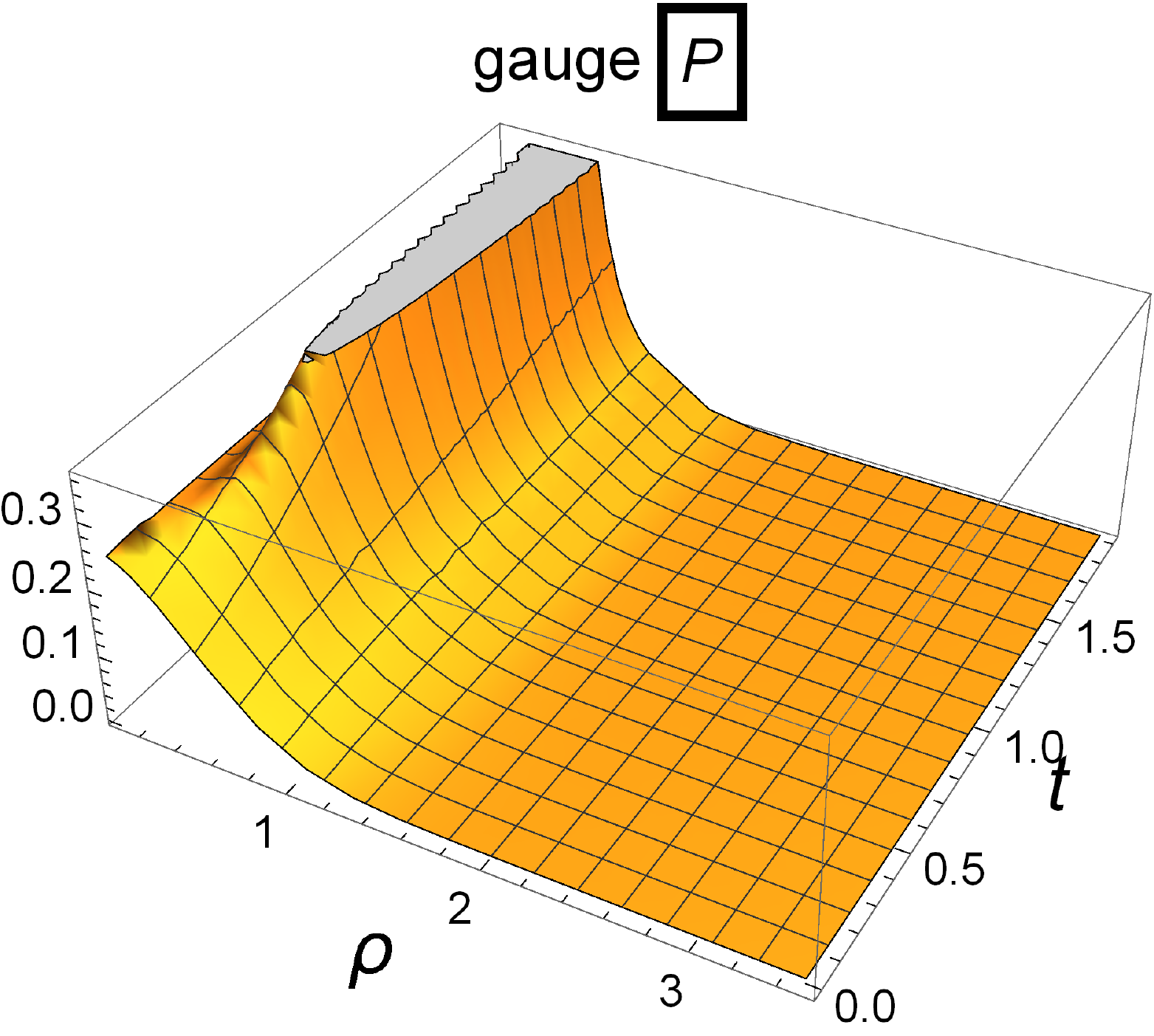}
\includegraphics[width=4.3cm]{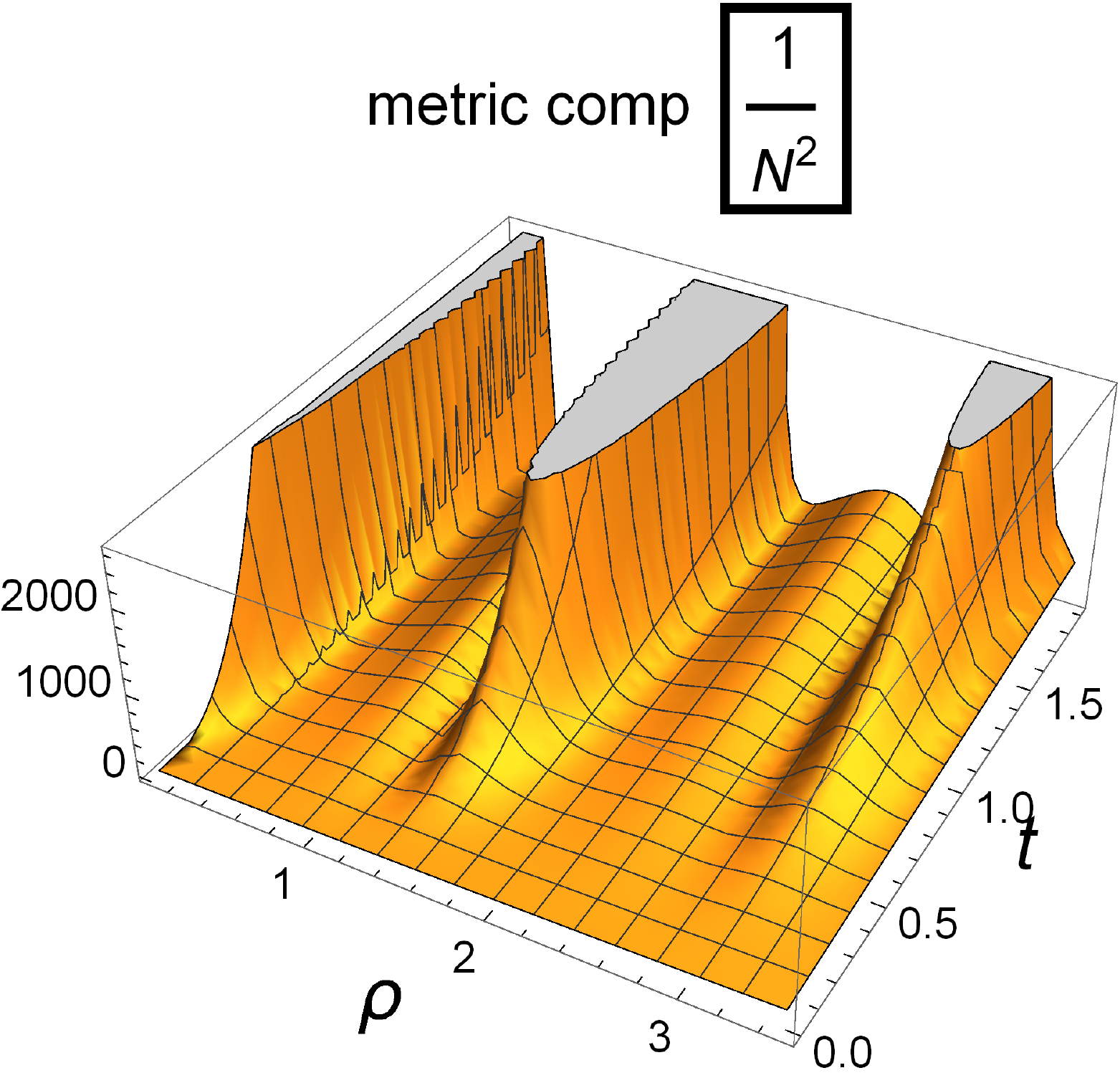}
\includegraphics[width=4.3cm]{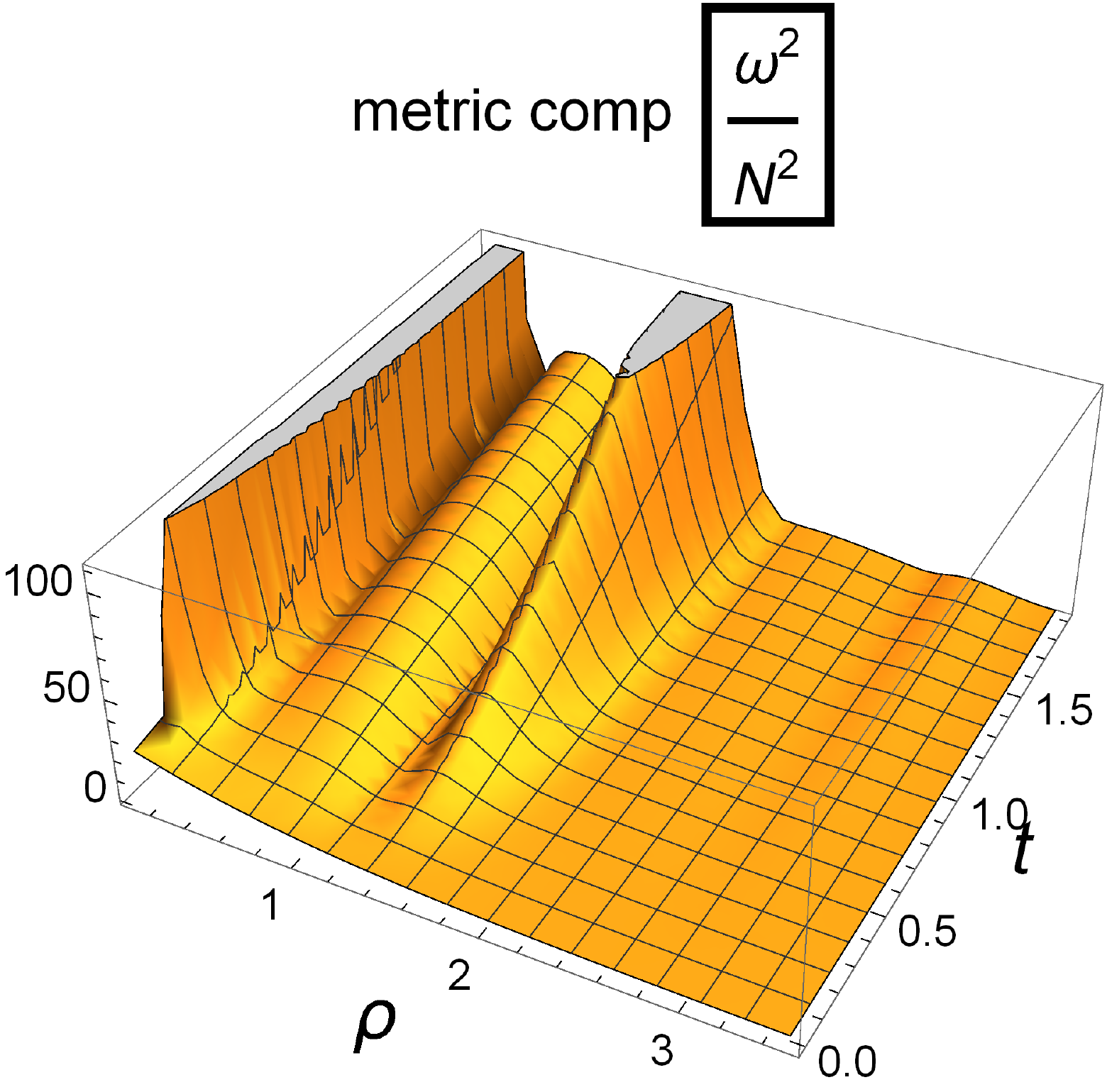}}
\centerline{
\includegraphics[width=4.3cm]{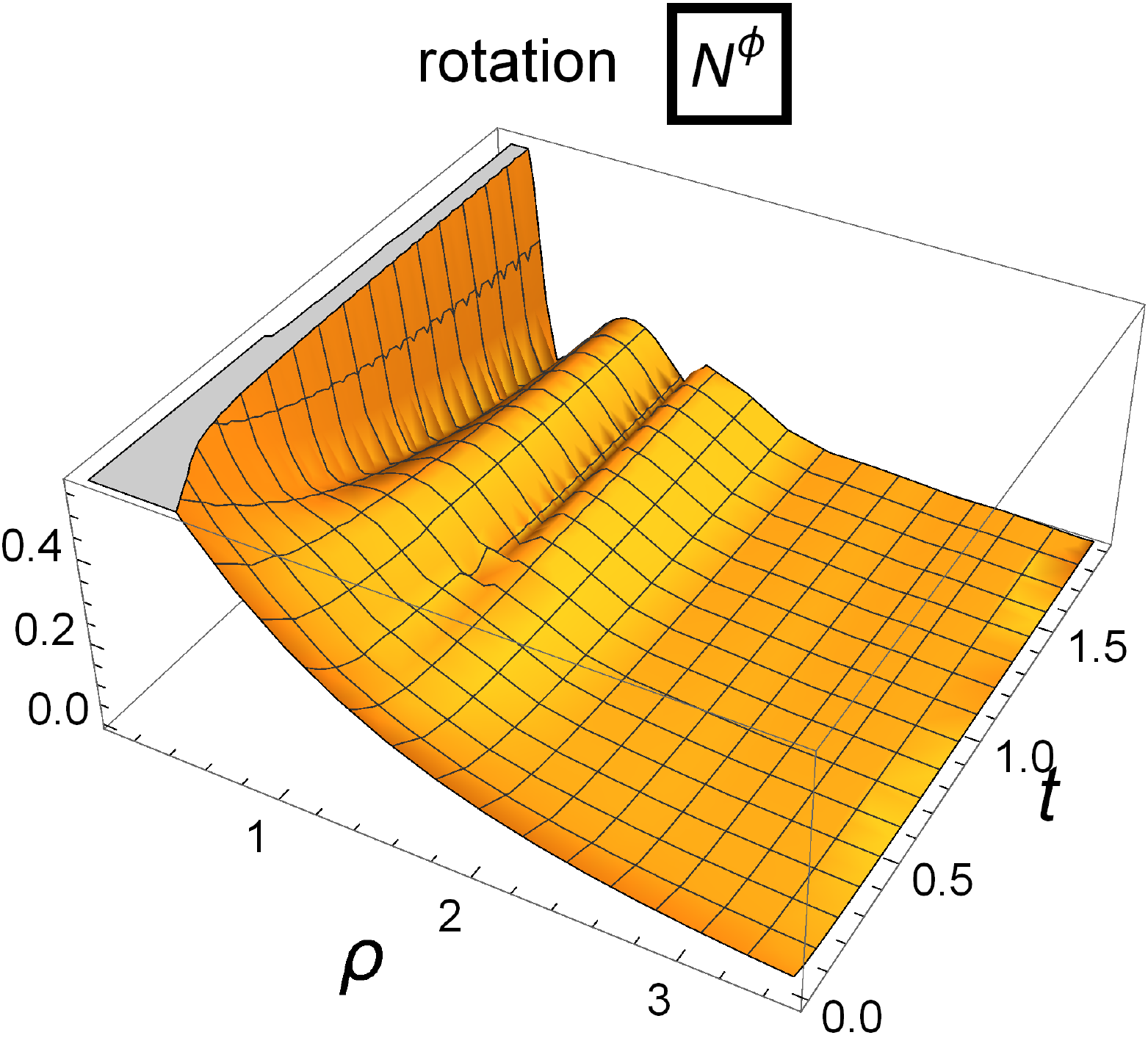}
\includegraphics[width=4.3cm]{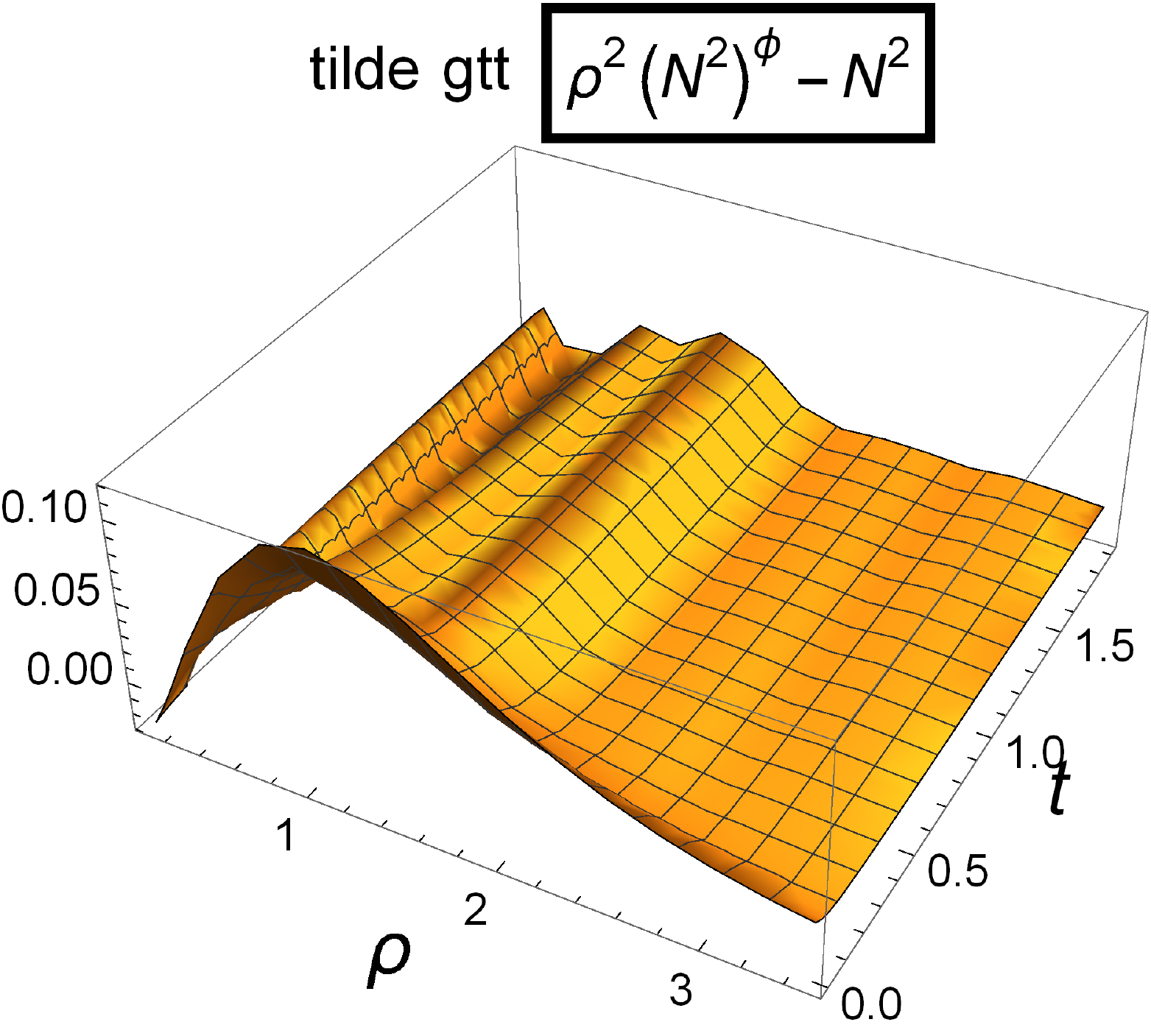}
\includegraphics[width=4.3cm]{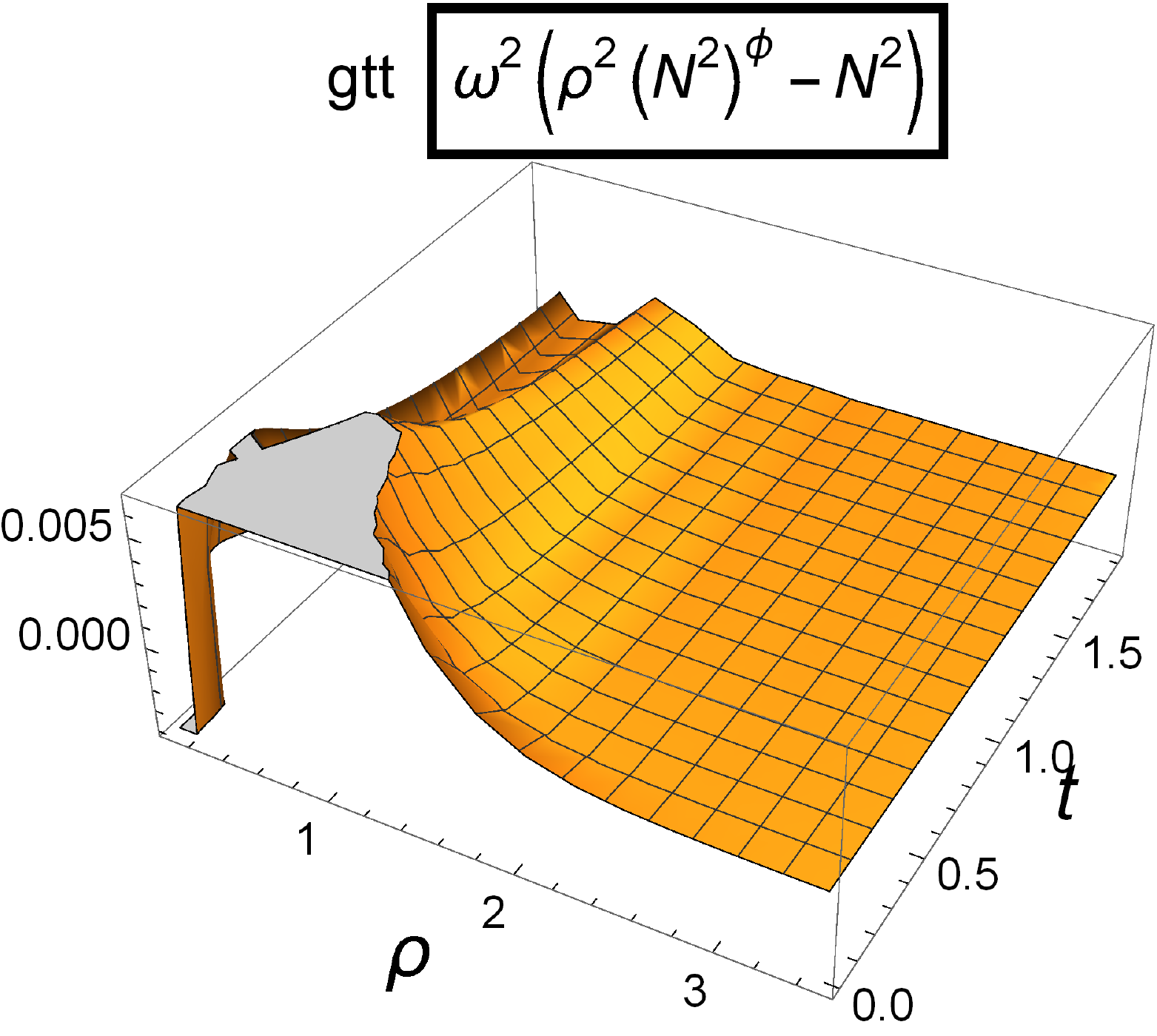}}
\vspace{1.cm}
\caption{{\it As figure 1, but now with a different initial wave for the scalar field. We used for the scalar field the initial outgoing wave $X=e^{-\rho}(\sin4(\rho-t)+\rho)$ and for the dilaton a decreasing mode. We observe that the wave turns quickly  into a solitary wave and induces a wavelike behavior in $\omega$.}  }
\end{figure}
\begin{eqnarray}
\omega=F_1(\rho)e^{-\frac{1}{2}k_1t}   ,\qquad   N=\frac{1}{F_1(\rho)}G(t-\log(\rho^{1/k_1}F_1(\rho)^{2/k_1}  ), \cr
 N^\varphi=H_1(t)+F_2(\rho)e^{k_1t},  \qquad F_2=a_1+a_2\int\frac{1}{\rho^3F_1^2}d\rho,\label{3-22}
\end{eqnarray}
where from the other PDE's a function for $F_1$ can be  found for any $k_1$.
In figure 2 and 3  we plotted  typical solutions for different initial and boundary conditions.
It turns out that the solution is insensitive for the cosmological constant (as expected), but very sensitive for the value of the potential.
Further, we observe that an initial wavelike function for the scalar field, {\it induces a wavelike behavior in the dilaton field}.
It is not a hard task to find the initial conditions and the suitable values of the several parameters in order to obtain a regular and singular free spacetime $g_{\mu\nu}$ out of a  BTZ solution $\tilde g_{\mu\nu}$ with it's horizon's.
For the vacuum flat situation ($N=1, N^\varphi =0$), we obtain for $\omega$ the Euler-Poisson-Darbois equation with solution
\begin{equation}
\omega =\frac{\alpha_1}{\rho^{\alpha_2}}e^{\alpha_3(t^2+\frac{1}{2}\rho^2)+\alpha_4t}.
\end{equation}
We already mentioned in the introduction, that the z-coordinate don't play a role in our model. So it was possible to uplift the BTZ-spacetime. We will return to this issue in connect with  {\it conformal compactification} in the next sections.
\section{An exact time-dependent vacuum BTZ solution in Eddington-Finkelstein coordinates in conformally invariant gravity}

Quite recently, some progress was made in understanding the physics at the horizon of black holes, where quantum effects will come into play. For a mainstream treatment on this subject we refer to Parker and Toms\cite{parker:2009}.

The fundamental question is what happens with an evaporating black hole (see for example the overview article of Page\cite{page:2004} and references therein). It is for sure that quantum effects will resolve the distinction
between the inside and outside of the black hole and the description of the hawking radiation.
It will be necessary to consider the {\it dynamical evolution} of the spacetime. This can be done in a tractable way in an Eddington-Finkelstein coordinate system.
One of the first attempts was the Vaidya solution\cite{vaidya:1943}. In fact, the Vaidya solution is one of the non-static solutions of the Einstein field equations and is a generalization of the static
Schwarzschild black hole solution. This solution is characterized by a dynamical mass function depending  on the retarded time. A solution on the (2+1)-dimensional BTZ spacetime was recently found by Chan, et al.\cite{chan:1994} and an approximate solution by Abdolrahimi, et al.\cite{abdol:2019}.
There are several ways to look at the "inside" of a black hole, or, differently formulated, how to extend maximally the Penrose diagram.  Some authors use the existence of white holes, a parallel universe, or  a wormhole to black-bounce transition\cite{simpson:2019,simpson:2019b}.
Another possibility  was proposed by Susskind and  Maldacena\cite{suss:2013}. Two entangled particles (a so-called Einstein-Podolsky-Rosen or EPR pair) are connected by a wormhole (or Einstein Rosen bridge) and may be a basis for unifying general relativity and quantum mechanics. However, the two entangled black holes in regions I and II in the  extended Penrose diagram,  will  interact via the ingoing and outgoing particles instantly.
Another problem is, how to treat the connection between the observation of the infalling observer and the outside observer, i.e., how to map the quantum states  of the in- and out-going radiation in a one-to-one  way.
In context of  conformally invariance and black hole complementarity, there is an other possibility of maximally extension of the Penrose diagram as initiated by {\it 't Hooft}\cite{thooft:2016}, using antipodal identification as spherical harmonics ( see also 't Hooft\cite{thooft:2019} and references therein).
If one don't want to give up locality and unitarity, one needs this approach.
We can ask ourselves if some of these ideas can be applied to our spacetime. It seems possible for the Kerr spacetime\cite{compere:2019}. However, here we are dealing with the uplifted standard (2+1) BTZ spacetime.
It is clear that one has to consider a dynamical evolution of the spacetime, as described  in section 3. In the case of the BTZ black hole, the {\it evolution of the horizons} (where the inner one is the instable Cauchy horizon) and ergo-surface outside the horizons can then be revealed.

Let us write the spacetime Eq.(\ref{3-1}) in the retarded ("outgoing") $U=t-\rho^*$ or advanced ("ingoing") $V=t+\rho^*$  Eddington-Finkelstein coordinates
\begin{eqnarray}
ds_r^2=\omega(U,\rho)^2\Bigl[-N(U,\rho)^2dU^2- 2dUd\rho+dz^2+\rho^2\Bigl(d\xi+N^\xi(U,\rho)dU\Bigr)^2\Bigr],\cr
ds_a^2=\omega(V,\rho)^2\Bigl[-N(V,\rho)^2dV^2+ 2dVd\rho+dz^2+\rho^2\Bigl(d\xi+N^\xi(V,\rho)dV\Bigr)^2\Bigr],\qquad\label{4-1}
\end{eqnarray}
with
\begin{eqnarray}
dU\equiv dt - \frac{d\rho}{N(t,\rho)^2} \equiv t-\rho^*,\qquad dV\equiv dt + \frac{d\rho}{N(t,\rho)^2} \equiv t+\rho^*, \cr
d\xi \equiv d\varphi -\frac{N^\varphi(t,\rho)}{N(t,\rho)^2}d\rho,\qquad\qquad\qquad\label{4-2}
\end{eqnarray}
with domains $U/V \in [-\infty ,+\infty ], \rho\in [-\infty ,+\infty ]$. We  make no a priori  assumptions for $N$ and $N^\xi$ (for example by writing $N=1-\frac{2M(u)}{\sqrt{\rho^2+a^2}}$\cite{simpson:2019}).
The field equations without matter terms now reduce to (an over-dot represents $\frac{\partial}{\partial U}$ and   $'=\frac{\partial}{\partial \rho}$)
\begin{eqnarray}
\omega''=\frac{2\omega'}{\omega},\label{4-3}
\end{eqnarray}
\begin{eqnarray}
2\omega N\omega'N'+3N^2\omega'^2+\frac{1}{\rho}\omega N^2\omega'+2\omega\dot\omega'+2\dot\omega\omega'+\frac{1}{\rho}\omega\dot\omega,\label{4-4}
\end{eqnarray}
\begin{eqnarray}
N''=-\frac{N'^2}{N}-\frac{4\omega' N'}{\omega}-\frac{3N\omega'^2}{\omega^2}-\frac{3N'}{r}-\frac{3N\omega'}{r\omega}+4\omega\dot\omega'-\frac{2\dot\omega\omega'}{N\omega^2}+\frac{3\dot\omega}{rN\omega},\qquad\label{4-5}
\end{eqnarray}
\begin{eqnarray}
4\dot\omega^2-2\omega\ddot\omega-\frac{\omega N^2\dot\omega}{r}+\frac{\dot NN\omega^2}{r}-6\omega'\dot\omega N^2-2\omega\dot\omega NN'+2\omega N\omega' N'+3{\omega'}^2 \cr
+2\omega\omega' N^3 N'+\frac{\omega\omega'N^4}{r}=0,\qquad\qquad\qquad\label{4-6}
\end{eqnarray}
\begin{eqnarray}
(\dot N^\xi)'=-2(N^\xi)'\frac{\dot\omega}{\omega},\qquad (N^\xi)''=-2(N^\xi)'(\frac{2\omega'}{\omega}+\frac{3}{r}),\label{4-7}
\end{eqnarray}
and constraint
\begin{equation}
(N^\xi)'^2=\frac{4}{3\rho^2}\Bigl[NN'+N'^2-3N^2\frac{\omega'^2}{\omega^2}-2N^2\frac{\omega'}{\rho\omega}+6\frac{\omega'\dot\omega}{\omega}+2\frac{\dot\omega}{\rho\omega}\Bigr].\label{4-8}
\end{equation}
One easily finds the non-trivial solution
\begin{eqnarray}
\omega=\frac{1}{e^{c_1 U}(c_2 \rho+c_3)},\qquad N^2=\pm c_1\frac{(c_3^2-c_2^2\rho^2)}{c_2c_3},\qquad N^\xi =F(U),\label{4-9}
\end{eqnarray}
with $F(U)$ an arbitrary function of $U$. This solution is consistent with the dilaton equation. Further, it is remarkable that the {\it time dependency emerge in} $\omega$ and not N.
However, $\tilde g_{UU}$ depends on U  via $N^\xi$.
So our metric $g_{\mu\nu}^{(4)}$ becomes (for the retarded case)
\begin{equation}
ds^2=\frac{e^{-2c_1U}}{(c_2\rho+c_3)^2}\Bigl[\pm\frac{c_1(c_3^2-c_2^2\rho^2)}{c_2c_3}dU^2 - 2dUd\rho+dz^2 + \rho^2\Bigl(d\xi+F(U)dU\Bigr)^2 \Bigr],\label{4-10}
\end{equation}
which is {\it flat}, while $\tilde R^{(4)}=\frac{6c_1c_2}{c_3}$. The function $F(U)$ will be fixed when matter terms are incorporated (i.e. for example, a scalar gauge field). The metric Eq.(\ref{4-1}) will then contain a term $b(U,\rho)^2d\varphi^2$ and a relation like
$(N^\xi)'=\frac{b}{\eta^2X^2+\omega^2}$ will be obtained.
We can now express, for example,  $U$  in $t$ and $\rho$:
\begin{equation}
U=t-\log\Bigl[\frac{c_2\rho +c_3}{c_2\rho -c_3}\Bigr].\label{4-11}
\end{equation}
So we have now a complete picture of the spacetime.
We must note that this solution is rather different with respect to the vacuum Vaidya spacetime. We also are dealing here with null radiation (null matter fields or gravitational radiation) as in the case of Vaidya, but we did not made any explicit assumption
for the $U$ or $V$ dependency of $\omega, N$ and $N^\xi$. They follow from the field equations. Further, the radiation is in the $(\rho,z)$-plane in stead of the $(r,\theta)$ plane in the Vaidya case. It is interesting to compare our solution with the Vaidya-type solution of a spinning  black hole in $(2+1)$ dimensions found by Chan in conventional GR\cite{chan:1994}. They also find a rotation function $N^\xi(U)$ which is determined by an energy-momentum tensor of null spinning dust.
From Eq.(\ref{4-10}), we see that the small-scale behavior (and so the dynamical apparent Cauchy horizon) is determined by
\begin{equation}
\omega^2(\rho^2N^{\xi 2}-N^2)=\frac{\rho^2F(U)^2 - \frac{c_1(c_3^2-c_2^2\rho^2)}{c_2c_3}}{e^{2c_1U}(c_2\rho+c_3)^2},\label{4-12}
\end{equation}
and in  the advanced coordinate
\begin{equation}
\omega^2(\rho^2N^{\xi 2}-N^2)=\frac{\rho^2F(V)^2 + \frac{c_1(c_3^2-c_2^2\rho^2)}{c_2c_3}}{e^{2c_1V}(c_2\rho+c_3)^2}.\label{4-13}
\end{equation}
If we omit the dilaton factor, we obtain the expressions for $\tilde g_{\mu\nu}$.
De apparent horizon is then determined (in V)  by
\begin{equation}
\frac{d\rho}{dV}=\frac{1}{2}(\rho^2(N^\xi)^2-N^2) =0,\label{4-14}
\end{equation}
so
\begin{equation}
\rho_{AH}=\pm\frac{c_3}{\sqrt{c_2(c_2-\frac{c_3}{c_1}F(V)^2)}},\label{4-15}
\end{equation}
and in U
\begin{equation}
\rho_{AH}=\pm\frac{c_3}{\sqrt{c_2(c_2+\frac{c_3}{c_1}F(U)^2)}}.\label{4-16}
\end{equation}
We see that the location of apparent horizon is independent of the dilaton (so also valid for $\tilde g_{\mu\nu}$). However, $g_{VV}$ depends also on $\omega$, as can be seen by inspection of Eq.(\ref{4-10}), i.e., the denominator.
\begin{figure}[h]
\centerline{
\includegraphics[width=5.5cm]{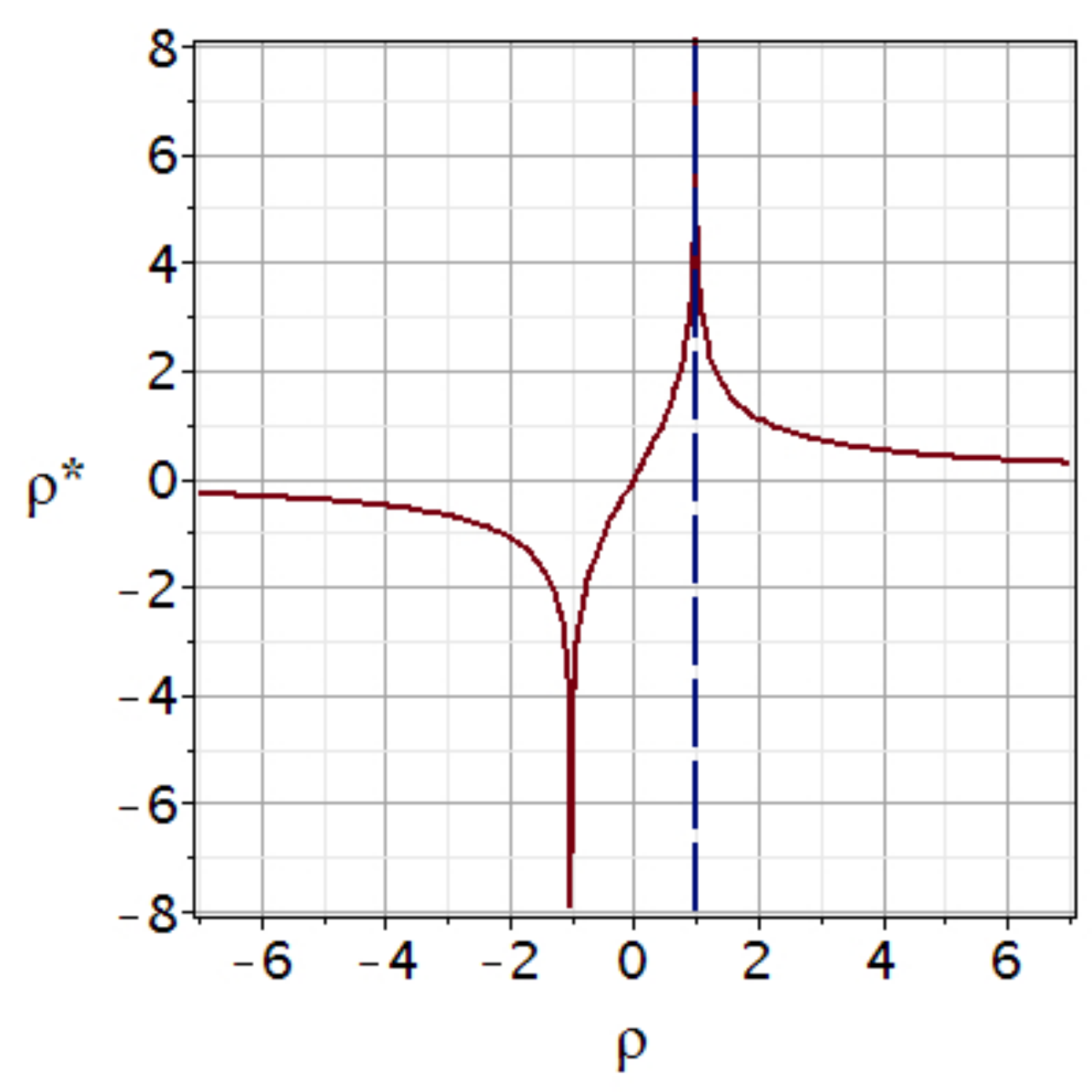}
\includegraphics[width=5.5cm]{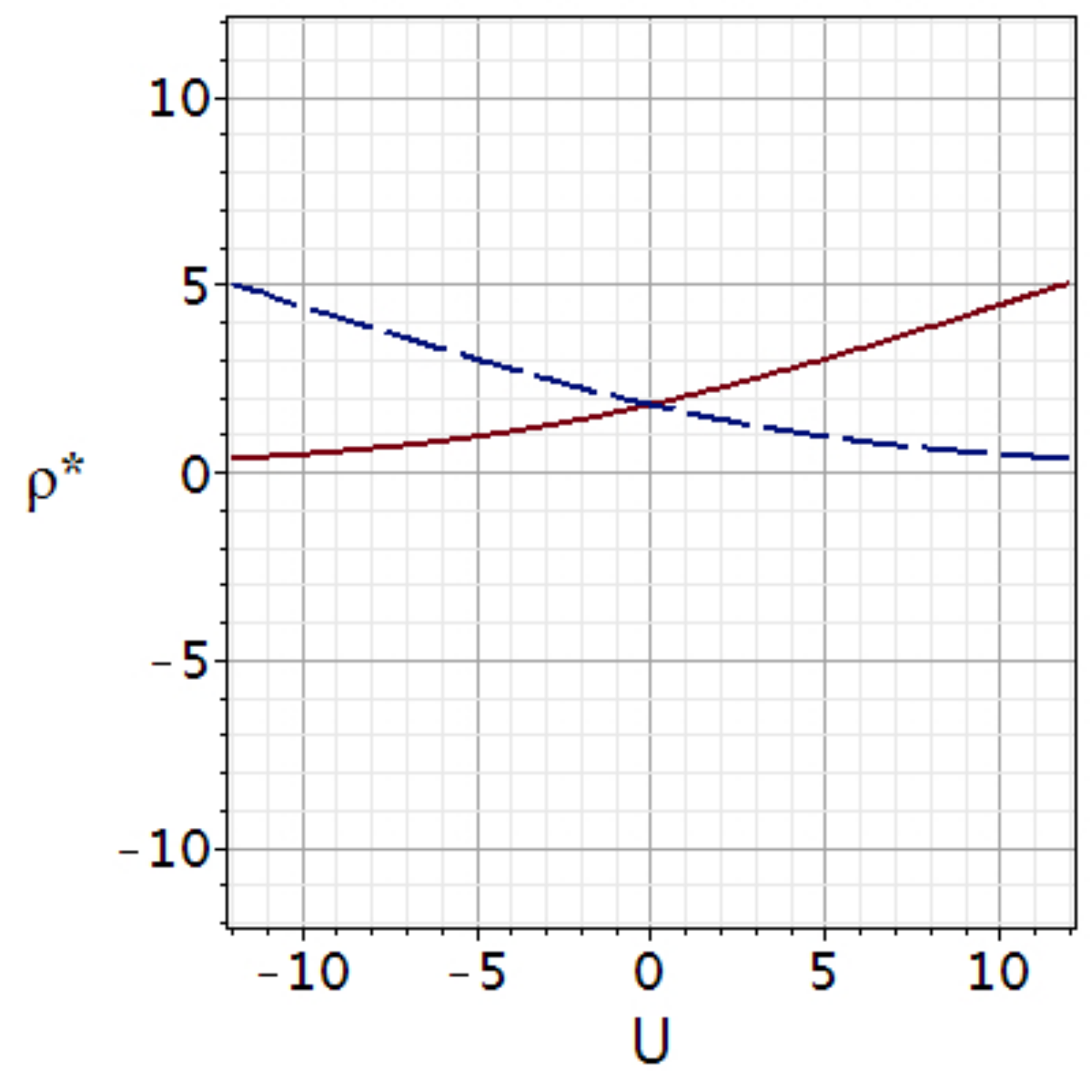}}
\caption{{\it Left: $\rho^*$ as function of $\rho$. The asymptote is at $\frac{c_3}{c_2}$. Right:  $\rho^*$ as function of U for $c_1=c_2=c_3=1$ and $c_4=\pm 0.3$.}}
\end{figure}
The solution turns out to be {\it also valid}   in the (2+1)-dimensional spacetime,
i.e., $\tilde R^{(3)}=\frac{6c_1c_2}{c_3}$ and $g_{\mu\nu}^{(3)}$ flat.
So we can safely uplift the BTZ solution in Eddington-Finkelstein coordinates in vacuum to 4-dimensional spacetime. We will return to this issue in the next section.
In figure 4 we plotted $\rho^*$ against $\rho$ and $\rho^*$ against $U$. The asymptote is at $\frac{c_3}{c_2}$.
In figure 5 we plotted the light cone structure. For the  outward emitted signals, the slope is given  by Eq.(\ref{4-14}) (in $U$) ( for the inward, $dU$=0).
For the limiting cases, we obtain
\begin{eqnarray}
\frac{d\rho}{dU}=\frac{1}{2}e^{-2c_1U}\cdot\begin{cases} -\frac{c_1}{c_2c_3}\:\:\:\ldots \ldots\ldots \rho \rightarrow 0 \cr
                           \frac{c_3F(U)^2+c_1c_2}{c_2^2 c_3} \:\:\:\ldots  \rho \rightarrow \infty \cr
                            \:\: \: 0 \:\:\:\:\ldots \ldots \ldots \ldots  \rho =\rho_{AH}\cr
                            \end{cases}
\end{eqnarray}
For $\rho \rightarrow 0$, its value doesn't tend necessarily to $-\infty$. For increasing $U$ it could approach zero again by suitable $c_1$.
Note that in general the location of the apparent horizon is dependent of $U$ (see figure 5).
We can express the apparent horizon also in $\rho^*$,
\begin{equation}
\rho_{AH}^*=\ln{\Bigl(\frac{1+\sqrt{1+\frac{c_3}{c_1c_2}F(U)^2}}{1-\sqrt{1+\frac{c_3}{c_1c_2}F(U)^2}}\Bigr)},\label{4-17}
\end{equation}
We can globally plot the location of the apparent horizon in a Penrose diagram, if we take for $F(U)$ for example $e^{c_4 U}$. See figure 5.

Let us now describe what is the meaning of the dilaton field for an infalling and outside observer in  connection with the complementarity  of the ingoing and outgoing massless particles. We will use the notion of  conformal maps.
The outside observed experiences a mass  $\omega^2N^2$  and an evaporating black  hole (in U-coordinate) by Hawking radiation (in the case of massive particles, of course, there can also be first a growing mass; we will not consider this here). This radiation is
\begin{equation}
\sim \partial_U(\omega^2 N^2)=\frac{2c_1^2}{c_2c_3}e^{-2c_1U}.\label{4-18}
\end{equation}
This blows up for $c_1<0$ and $U\rightarrow  +\infty.$
However, there is in $g_{VV}$ in the denominator the factor $e^{2c_1U}$. So an infalling observer crossing the apparent horizon will need a different $\omega$. The ingoing observer, passing the horizon, will NOT use the $\omega$ of the outside observer.  In fact, it is locally unobservable. This happens at very small scales, when $g_{UU}\rightarrow 0$ and $\omega^2(\rho^2N^{\xi 2}-N^2)\rightarrow 0$
for $U<<-1$ in Planck units (the ergo-surface) and there is no horizon at all (note that $\omega^2$ is an overall factor for $\tilde g_{\mu\nu}$).
The dilaton determines the different notion of what is happening near the horizon for an infalling and outside observer.
Now remember that the Ricci scalar curvature transforms under conformal transformations as $R\rightarrow\frac{1}{\Omega^2}\Bigl(R-\frac{6}{\Omega}\nabla^\alpha\nabla_\alpha\Omega\Bigr)$ and the additional freedom in $\omega$, i.e., $\omega \rightarrow\frac{1}{\Omega}\omega$. The dilaton equation of  Eq(\ref{3-5}) is an auxiliary  equation in vacuum. It follows also from the Einstein equations. When matter is included, one obtains conditions on the potential (see, for example, Eq.(\ref{3-18})).  So it would be fine if we could impose $R=0$ for the local observer by using $\tilde R-\frac{6}{\Omega}\tilde \nabla^\alpha\tilde \nabla_\alpha\Omega =0$.
One can then apply Fourier analysis of  quantum mechanics and treat $\omega$ infinitesimal\cite{thooft:2009}. This is a complementarity transformation on the dilaton and {\it switches on and off} the effects these Hawking particles have on the metric.

\begin{figure}
\begin{tikzpicture}
  \begin{axis}[domain  = -5:5,
               samples = 100,
               xmin    = -3,
               xmax    = 5,
               ymin    = -3,
               ymax    = 3,
               ytick   = \empty,
               xtick   = \empty,
               xlabel near ticks,
               ylabel near ticks,
               set layers,              ]
   \node[anchor= north] at (axis cs: 3.42,1) {$\frac{c_3}{c_2}\Bigl(\frac{1}{\sqrt{1+\frac{c_3}{c_1c_2}F(U)^2}}\Bigr)$};
   \node[anchor= north] at (axis cs: 4.6,-0.15) {$\rho$};
  \node[anchor= north] at (axis cs: -0.3,3) {$U$};
    \draw[black,thick] (axis cs: -5, 0 )-- (axis cs: 5, 0);
    \draw[black,thick] (axis cs: -1, -3 )-- (axis cs: -1, 3);
     \draw[red,dashed,thick] (axis cs: 1.8, -3 )-- (axis cs: 1.8, 3);
      \draw[blue,thin] (axis cs: -2, -2. )-- (axis cs: 5, -2.);

\end{axis}
\draw[darkgreen,thick,fill=green, fill opacity=0.4] (2,1.18) circle [x radius=0.248cm, y radius=1.mm, rotate=60];
\draw[darkgreen,thick,fill=green, fill opacity=0.4] (2.6,0.73) circle [x radius=0.248cm, y radius=1.mm, rotate=60];

\draw[darkgreen,thick]  (2.47,0.54) -- (2.16,1.36);
\draw[darkgreen,thick]  (1.86,0.95) -- (2.7,.95);


\draw[darkgreen,thick,fill=green, fill opacity=0.4] (3.91,1.19) circle [x radius=0.3cm, y radius=1.mm, rotate=45];
\draw[darkgreen,thick,fill=green, fill opacity=0.4] (4.33,0.73) circle [x radius=0.3cm, y radius=1.mm, rotate=45];

\draw[darkgreen,thick]  (4.12,1.4) -- (4.12,0.59);
\draw[darkgreen,thick]  (3.73,0.95) -- (4.53,0.95);


\draw[darkgreen,thick,fill=green, fill opacity=0.4] (6.2,1.175) circle [x radius=0.36cm, y radius=1.mm, rotate=30];
\draw[darkgreen,thick,fill=green, fill opacity=0.4] (6.45,0.72) circle [x radius=0.36cm, y radius=1.mm, rotate=30];

\draw[darkgreen,thick]  (6.5,1.3) -- (6.15,0.55);
\draw[darkgreen,thick]  (5.96,0.95) -- (6.7,0.95);

\draw[->,red,thick] (4.1,0.5) to [out=100,in=280] (3,4);
\end{tikzpicture}
\hspace{2cm}
\begin{tikzpicture}
  \begin{axis}[domain  = -8:8,
               samples = 100,
               xmin    = -9,
               xmax    = 9,
               ymin    = -8.5,
               ymax    = 8.5,
               ytick   = \empty,
               xtick   = \empty,
               xlabel near ticks,
               ylabel near ticks,
               set layers,
              ]

    \draw[black,dashed,thick] (axis cs: -8, 0 )-- (axis cs: 8, 0);
    \draw[black,dashed,thick] (axis cs: 0, -8 )-- (axis cs: 0, 8);

     \node[anchor= north] at (axis cs: 5,6) {$V$};
     \node[anchor= north] at (axis cs: -5,6) {$U$};
     \node[anchor= north] at (axis cs: 8.3,0) {$\iota^0$};
      \node[anchor= north] at (axis cs: 1.,8.8) {$\iota^+$};
     \node[anchor= north] at (axis cs: 6.5,1.2) {$t$};
     \node[anchor= north] at (axis cs: 0.8,6.5) {$\rho^*$};
\node[anchor= north,red] at (axis cs: 4,3) {$AH$};
\node[anchor= north,green] at (axis cs: -1.7,4.9) {$AH$};
     \draw[blue,thick] (axis cs: -5, -5 )-- (axis cs: 4, 4);
     \draw[blue,thick] (axis cs: 4, -4 )-- (axis cs: -4, 4);
     \draw[blue,thick] (axis cs: 0, 8 )-- (axis cs: 8, 0);
     \draw[blue,thick] (axis cs: 0, -8 )-- (axis cs: 8, 0);
     \draw[blue,thick] (axis cs: -8., 0 )-- (axis cs: 0, 8);
\end{axis}
\draw[->,red,thick] (4.7,3.5) to [out=227,in=310] (2.45,3.93);
\draw[->,green,thick] (2.9,4.9) to [out=275,in=135] (4.76,1.8);

\draw[->,red,thick,dashed] (2.9,1.65) to [out=71,in=305] (2.38,3.65);
\draw[->,green,thick,dashed] (1.4,3.5) to [out=340,in=135] (4.48,1.65);

\end{tikzpicture}
\caption{{\it Left: Plot of light cone structure. The location of the apparent horizon is for small $F(U)$ at $\frac{c_3}{c_2}$. For increasing $F(U)$ it tends closer to $\rho =0$. Right: Penrose diagram for the evaporating BTZ black hole in Eddington Finkelstein coordinates $(U,\rho^*)$. The global location of the two pairs of  the apparent horizons as function of $U$  are indicated. Note that one pair  $\rho^*_{AH}$  enters the "future inside" region and comes from the "past inside". By the antipodal identification these  regions are removed ( no "inside" of the black hole) so the locations are  mapped on each other. See section 5.  }}
\end{figure}
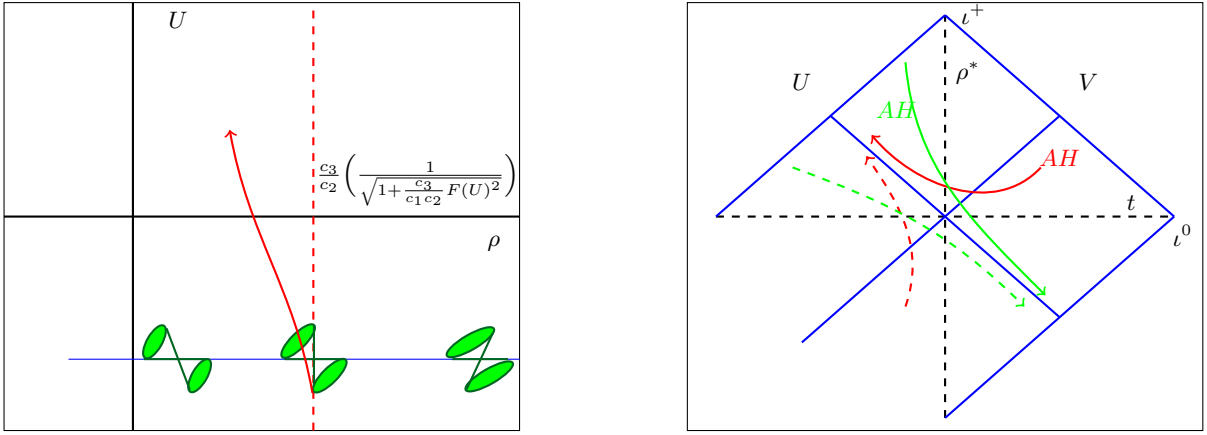
\section{Complementarity transformation and conformal compactification}
Let us return to the {\it conformal mapping} in more detail. We know that in Minkowski spacetime the conformal map preserves the light-cone structure and so the null geodesics (i.e., the affine parameter). The conformal group in Minkowski, however, does not act as linear transformations, so one needs a trick (see, for example Felsager\cite{felsager:1998}, section 10.3 and figure 6.).
\begin{figure}[h]
\centerline{
\includegraphics[width=8.0cm]{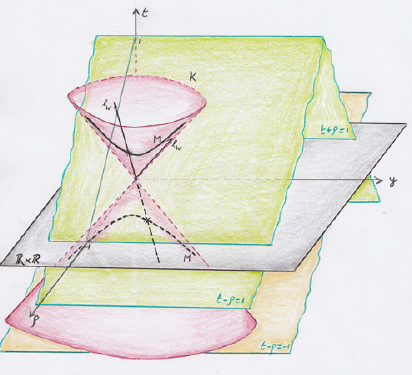}
\includegraphics[width=4.5cm]{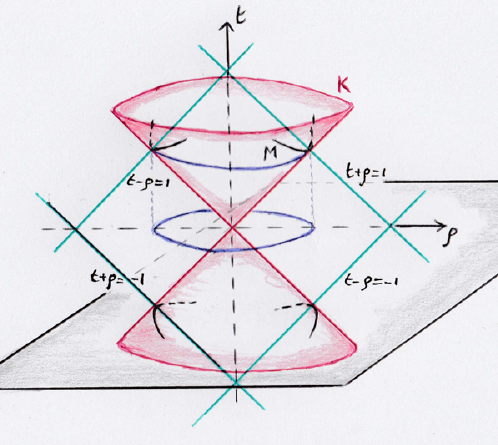}
\includegraphics[width=4.cm]{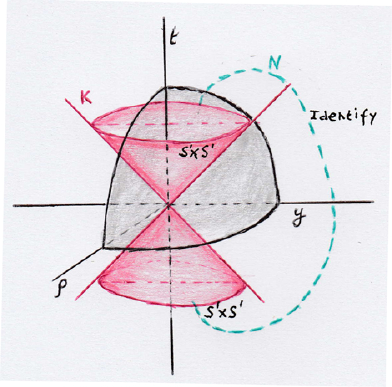}}
\caption{{\it The conformal compactification. We  have visualized the hyperplanes $t-\rho=\pm 1$ and null cone $K$. For completeness, we also visualized the plane $t+\rho =1$. The $z$-coordinate is suppressed. The horizontal plane is the pseudo-Cartesian space ${\bf R}^1 \otimes {\bf R}^1$. $M$ is the intersection of the hyperplanes with the null cone $K$. A generator of $K$ is a null vector and generates characteristic lines
$\ell_w=\tau(t,\rho ,z,y)$, which intersects $M$ 0 or 1 time (or more then 1 by a conformal map in the region $t<0$). The null vectors $y^2+z^2=0$ ($z$ suppressed) parallel to the hyper-planes intersect $K$ in the origin, which represents  a point at infinity in the enlarged space of the sphere by stereographic projection (see text).
The enlarged space $M({\bf R}^2 \otimes {\bf R}^2)$ is topologically equivalent with ${\bf S}^1 \otimes {\bf S}^1$. If $N$ represents the intersection of $K$ with the hyper-sphere $t^2+\rho^2+y^2=2$, then $N$ is topologically a hyper-torus (because the z-coordinated was suppressed) and a characteristic line will intersect ${\bf S}^1 \otimes {\bf S}^1$ is two antipodal points. Note that time-inversion take place, which means in quantum mechanical language that the creation and annihilation operators are interchanged\cite{thooft:2016}.
We obtain the inverted $M$ after the intersection of $K$ with the planes $t=\rho =\pm 1$.
$M({\bf R}^2 \otimes {\bf R}^2) \rightarrow {\bf S}^1 \otimes {\bf S}^1$ is a conformal map\cite{felsager:1998}.
${\bf S}^1 \otimes {\bf S}^1$, a compact subset of ${\bf R}^2 \otimes {\bf R}^2$, is obtained from ${\bf R}^1 \otimes {\bf R}^1$ by adding a cone at infinity. This is a conformal compactification of ${\bf R}^1 \otimes {\bf R}^1$. }}
\end{figure}
One starts with a pseudo-Cartesian space ${\bf R}^1 \otimes {\bf R}^1$ ( for example our $(x,z)$). One then enlarge first the pseudo-Cartesian space by adding  a "null"-cone at infinity. So one compactifies the plane
in ${\bf R}^2$. In order to apply the conformal transformation of inversion, one considers the unit sphere ${\bf S}^1$ and map ${\bf R}^1$ onto ${\bf S}^1-\{N\}$ (one sends the north pole to infinity).
If we want to apply all the conformal transformations, then we must enlarge the pseudo-Cartesian space by adding two extra dimensions $(t,y)$, (later, we replace x by $x=\rho\sin\varphi$ and $y=\rho\cos\varphi$, to get back our axially symmetric spacetime coordinates $(t,\rho,z,\varphi)$). The goal is then to embed the pseudo-Cartesian space ${\bf R}^1 \otimes {\bf R}^1$ as a subset of ${\bf R}^2 \otimes {\bf R}^2$.
We define $M$ as the intersection of the null cone K (in  ${\bf R}^2 \otimes {\bf R}^2$) with the hyperplane $\rho-t=1$ (or $\rho+t$) and define an isometry $ F: {\bf R}^n \otimes {\bf R}^n \rightarrow M$. Further, one works in the particular section of $M$, $M({\bf R}^1 \otimes {\bf R}^1)$. Because $\cal{F}$ induces a coordinate system on $M$, one can construct characteristic lines $\ell_w$. There are characteristic lines that are parallel to $\rho-t=1$ and are generated by null vectors where $\rho =t$. There is a one-to-one correspondence between these lines missing $M({\bf R}^1 \otimes {\bf R}^1)$ and points on $K$ in  ${\bf R}^1 \otimes {\bf R}^1$. So they represents points on the null cone at infinity. See figure 4.
\begin{wrapfigure}{c}{0.5\linewidth}
\centering
\includegraphics[width=6.0cm]{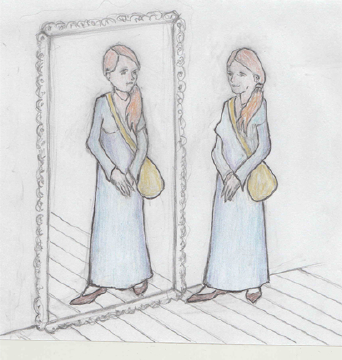}
\caption{{\it Antipodal map: there is a parity reflection after $z \rightarrow -z$, $\varphi \rightarrow \varphi +\pi$ [drawing:  Titia Oppenhuis de Jong]}}
\end{wrapfigure}
One can proof\cite{felsager:1998} that local sections $N_1$ and $N_2$ on the null cone which intersect characteristic lines at most once, can be mapped onto each other by a conformal map obtained by projection along the characteristic lines.
If we would now try to project $M({\bf R}^1 \otimes {\bf R}^1)$ onto a suitable subsection of $K$, then it turns out that it is not possible to find a single section that is intersected exactly once by each characteristic line.
Instead  one can consider N as the product of two unit spheres in $M({\bf R}^2 \otimes {\bf R}^2)$, i.e., N becomes a hyper-torus ${\bf S}^1 \otimes {\bf S}^1$ and each characteristic line will then intersect $K$ twice in antipodal points. So each point in $M({\bf R}^1 \otimes {\bf R}^1)$ is represented by a pair of  antipodal points on ${\bf S}^1 \otimes {\bf S}^1$. The projection is a conformal map.
The procedure here described is called a conformal compactification of $M({\bf R}^1 \otimes {\bf R}^1)$.
In our case the antipodal identification is $(U,V,z,\varphi)\rightarrow (-U,-V,-z,\varphi +\pi)$ (or $y=\rho\cos\varphi \rightarrow -y$). The points are ${\it not}$ physically distinct events, but identical and are different representations of one black hole. In fact, there is no inside of the black hole.
The price is that the manifold is {\it not time-orientable} for $\rho < \rho_{AH}$ (but one avoids the existence of a firewall when considering quantum mechanical issues).

One can see the antipodal identification as a "boundary" condition: it removes the "inside" of the black hole.  Nothing ever escapes to the interior. For the local observer the effect is invisible (see figure 6).

Note that time-inversion must take place when crossing the horizon (topological a M\"{o}bius strip\footnote{Note that we could also make a "Klein"-bottle identification, in analogy with the M\"{o}bius construction  in 2 dimensions. However, it can only be realized in 4D space dimensions, which means that we need the warped 5D spacetime\cite{slagter:2016}.}), which means in quantum mechanical language that the creation and annihilation operators are interchanged\cite{thooft:2016}.
But the antipodal identification is  symmetric under CPT: the outside observer will still experience CPT invariance
In figure 6 we visualized the construction of the hyper-torus.

When the evaporation process speeds up, we observe  from Eq.(\ref{4-15}), (\ref{4-16}) and (\ref{4-17})  that the two horizons approach $\rho^*$ is zero for increasing $F(U)$, which is assumable. Moreover
\begin{equation}
\lim_{\rho\rightarrow 0} g_{UU}\rightarrow \pm \frac{c_1}{c_2c_3e^{2c_1U}},\label{5-2}
\end{equation}
where in the denominator appears the exponential factor from the dilaton. So $\omega$ determines the scale as function of $U$ the local observer experiences. Note that on {\it "the other side"} ( in the  Penrose picture region II), $U$ change sign and the righthand side of Eq.(\ref{5-2}) becomes $\frac{c_1}{c_2c_3e^{-2c_1U}}$.

We found in section 4 that the solution of the BTZ spacetime in 4D in Eddington-Finkelstein  coordinates in conformally invariant gravity is identical to the 3D case, where we omitted the $dz^2$.
That is curious, because we can still apply the conformal compactification (conformal transformations) and the antipodal identifications in 4D spacetime sketched above.
Further, we obtained a {\it flat} $g_{\mu\nu}$ out of the "un-physical" $\tilde g_{\mu\nu}$, which resembles the original BTZ-black hole (without the need of a cosmological constant).
Some  notes can be made about the connection with the gravitational back-reaction. In the non-vacuum situation of section 3, the back-reaction is quite clear. In the vacuum case, there will be  a shift in the location of the apparent horizon after the emission of null radiation ( Eq.(\ref{4-18})). This can be made clear in the Penrose diagram,  as was also found in the time-dependent Vaidya spacetime in connection with black-bounces and traversable wormholes\cite{simpson:2019,simpson:2019b}. In the conformally invariant model and the antipodal approach,  however, one don't need such extreme escape.  Moreover, these two regions will communicate to one another quantum mechanically, which violate unitarity.
This shift will be related to the $c_i$  in  Eq.(\ref{4-17}), just as the scalar curvature of $\tilde g_{\mu\nu}$ was related to $c_i$, i.e., $\tilde R^{(4)}=\frac{6c_1c_2}{c_3}$ . A comparable effect was found in the counterpart model of the cylindrical radiating Lewis-van Stockum solution (in ($\rho,t)$ coordinates) and Einstein-Rosen pulse-wave solution\cite{islam:1985}. This solution is obtained from the stationary  $(z,\rho)$ spacetime where one replaces
$t\rightarrow iz,  z \rightarrow it$ and $ J\rightarrow iJ$. This solution has, however, reflection symmetry, $\varphi \rightarrow -\varphi, z\rightarrow -z$. A curious feature of the solution is the fact that the Kretschmann scalar becomes zero for two different values of a constant in the exact solution. In some sense, the spacetime returns to his original status after the emission of the pulse wave.
The relation with the antipodal symmetry is current under investigation by the author.

\section{Connection with the M\"{o}bius transformations}
In language of group theory, the map of a pair of antipodal points into a pair of antipodal points can be considered as a conformal transformation on $M({\bf R}^1 \otimes {\bf R}^1)$ and represented by the pseudo-orthogonal group of matrices $O(2,2)$. The matrix $-\mathbb{I}$ will interchange antipodal points.
However, the correspondence is not one-to-one. A point in ${\bf R}^1\otimes {\bf R}^1)$ corresponds to a pair antipodal points on ${\bf S}^1 \otimes {\bf S}^1$.
The conformal transformations generated from $O(2,2)$ form the conformal group $C(1,1)$. Each conformal transformation in $C(1,1)$ can then be represented by a pair of antipodal matrices in $O(2,2)$.
Just as in the construction of the conformal compactification, by adding the a "cone at infinity" (${\bf S}^1 \otimes {\bf S}^1$ from ${\bf R}^1 \otimes {\bf R}^1$) , we need not double count the points $\{+1\} {\bf S} \otimes \{-1\}\otimes{\bf S}$. So we simple throw away the $\{1\}\otimes {\bf S}$. A pseudo orthogonal matrix that interchange antipodal points will therefore generate the identity, $-{\bf I}$.
The conformal group $C(1,1)$ constitutes of  inversions, translations, dilatations and special conformal translations.
These matrices $MC$ in $O(2,2)$ generate conformal transformations on ${\bf S}^1\otimes {\bf S}^1$, $F(CM)$, preserving the inner product in ${\bf R}^2\otimes {\bf R}^2$ and maps the null cone into itself.
One can proof that $F(CM)$ are the conformal transformations of the hyper torus into itself. It maps a pair of antipodal points into a pair of antipodal points and can also be considered as conformal transformations on ${\bf R}^1\otimes {\bf R}^1$.

The construction of the conformal compactification is often represented by the M\"{o}bius transformations of the complex plane onto itself, i.e., $M({\cal C}_\infty)=\{ f:\mathbb{C}_\infty\rightarrow\mathbb{C}_\infty|f(z)=\frac{az+b}{cz+d}, ad-bc\neq 0\}$.
The M\"{o}bius transformations are one-to-one analytic (complex differential) maps of the unit disk to itself\cite{olsen:2010}.
The stereographic projection with an one point compactification (null cone to infinity) on the complex plane, SP: $S^2-\{N\}\rightarrow {\bf C}^1$ ( and its inverse) is bijective, conformal and continuous, so a homeomorphism.
Equipped with this complex coordinate system, the sphere ${\bf S}^2$ is known as the {\it Riemann sphere}.
The matrices can also be classified as a group under composition of functions (surjective): $\Gamma: GL_2(\mathbb{C})\rightarrow M(\mathbb{C}_\infty)$, with kernel $Ker(\Gamma)=k{\bf I}$. If the determinant is 1, $k=\pm1$ (the group $PGL_2$).

The  M\"{o}bius transformations also preserve orthogonality and symmetry (two points are symmetric with respect to $C$ if and only if their images under $F$ are symmetric with respect to the image of $C$).

We now know that there can be no fixed points ($f(z)=z$). This can be formulated nicely using the $GL_2(\mathbb{C})$ as rotations. Rotations of $\mathbb{C}_\infty$ are  M\"{o}bius transformations. A rotation has  an axis that is fixed. So if we choose a suitable orthonormal basis, with the vector that is fixed as a basis vector, of ${\bf R}^3$, we can write the matrix
\begin{equation*}
A =
\begin{pmatrix}
\cos\theta & \sin\theta & 0 \\
-\sin\theta & \cos\theta & 0 \\
0 & 0 & 1
\end{pmatrix}
\end{equation*}

A rotation of ${\bf S}^2$ extends to a rotation of ${\bf R}^3$ by linearity (=$SO(3)$). The rotations are conformal. Formally we have, $f: {\bf C}_\infty \rightarrow{\bf C}_\infty$ is a rotation of ${\bf C}_\infty$ if the map
$SP^{-1}\circ f\circ  SP: {\bf S}^2 \rightarrow {\bf S}^2$ is an element of $SO(3)$.
If $P=(U,V,$z$,\varphi)\in {\bf S}^2$ (where we  suppress z\footnote{note the difference here between z and $z$}) with antipodal point $\tilde P(-U,-V,$-z$,\varphi +\pi)$ and if $z=SP[(U,V,\varphi)]\in {\bf C}_\infty$, then the antipodal point $\tilde z \in {\bf C}_\infty$ is given by $\tilde z =SP[(-U,-V,\varphi +\pi)]$. From the features of the stereographic projection one then obtains that the antipodal point of $z$ is given by $\frac{-1}{\bar z}$ and $f(\frac{-1}{\bar z})=\frac{-1}{\bar f(z)}$ (sometimes named as inversions). So if $f\in Rot({\bf C}_\infty$, then antipodal pairs $(z,\tilde z)$ are mapped to antipodal pairs $[f(z),f(\tilde z)]$. Since rotating ${\bf S}^2$ and then apply the antipodal map is the same as applying the antipodal map and then rotating.
There is a group isomorphism $Rot({\bf C}_\infty \cong SO(3)$.
If the M\"{o}bius transformations is a translation, then $\infty$ is the only fixed point. If the M\"{o}bius transformations is a dilation, then 0 and $\infty$ are the only fixed points.
Further, one can proof that an element in $PGL_2(\bf C)$ whose fixed points are  antipodal in ${\bf C}_\infty$  is a rotation of ${\bf C}_\infty$.

\section{Conclusion}
Using  conformally invariant gravity, a new solution is found for the uplifted BTZ spacetime, without a cosmological constant. The solution shows some different features with respect to the standard BTZ solution.
In the non-vacuum situation, where a scalar-gauge field is present, a numerical solution is presented on a spacetime where one writes the metric as $g_{\mu\nu}=\omega^2 \tilde g_{\mu\nu}$, with $\omega$ a dilaton field, to be treated on equal footing with the scalar field and $\tilde g_{\mu\nu}$ an "un-physical" spacetime. The effect of $\omega$ on the behavior of the solution is evident. An outgoing wave-like initial value for the scalar field induces a wave-like response in the dilaton field and pushes the apparent horizon closer to $\rho =0$. The solution depends critically on the shape  of the potential. The solution can be used to investigate what happens with the spacetime of an evaporating black hole through  Hawking radiation.
In the vacuum situation in Eddington-Finkelstein coordinates, an exact solution is found for the (2+1)-dimensional case {\it as well as} for the uplifted situation. The "un-physical" $\tilde g_{\mu\nu}$ (BTZ) solution has a non-zero Ricci scalar, while  $g_{\mu\nu}$ {\it is flat}.

There is possibly a link with the antipodal identification. Antipodal mapping is inevitable if one wants maintain unitarity during quantum mechanical calculations on the Hawking particles.
The antipodal identification can be represented as a conformal transformation generated from the pseudo-orthogonal matrices of $O(3)$, i.e., the {\it conformal group}. Each conformal transformation in this group can be presented by a pair of antipodal matrices.
This was the main reason to investigate in this manuscript the dynamics of the BTZ black hole in conformally invariant gravity.
In the conformally invariant approach plays the dilaton field a fundamental role. We find that as soon as its value is fixed (by the global spacetime after choosing the coordinate frame), the local observer experiences scales.
Moreover,we find that  it also plays a role in the {\it antipodal mapping}. If we substitute the apparent horizon $\rho_{AH}$ (Eq.(\ref{4-16})) into $\omega$ (Eq.(\ref{4-9})) at the horizon, we can then compare $\omega$ on both sides of the horizon by replacing $U$ by $-U$. By imposing proper matching conditions, one could obtain restrictions  on $F(U)$.

We do not pretend that out model is a new  description of the physics of an evaporating BTZ black hole. We have tried to compare  conformally invariant gravity solutions of the (2+1)-dimensional BTZ black hole solution and its uplifted counterpart model with the results of former results on black hole studies. Specially the antipodal identification seems to fit well in our model.

\section{References}
\end{document}